\documentstyle[preprint,floats,prb,eqsecnum,aps]{revtex}
\input epsf
\begin{document}

\date{29$^{\rm th}$ November 1994}
\title{Topological Defects in the Abrikosov Lattice of Vortices in Type-II
Superconductors}
\author{M.J.W. Dodgson and M.A. Moore}
\address{Theoretical Physics Group \\
Department of Physics and Astronomy \\
The University of Manchester, M13 9PL, UK}
\maketitle
\draft 
\tighten

\begin{abstract}

The free energy costs for various defects within an Abrikosov lattice of 
vortices  are calculated using the lowest Landau level approximation (LLL).
Defect solutions with boundary conditions for lines to meet at a point
(crossing defect) and for lines to twist around each other (braid defect) are
sought for 2, 3, 6, and 12 lines. 
 Many results have been unexpected, including the nonexistence
of a stable two- or three-line braid. This,
 and the high energy cost found for a six-line
braid lead us to propose that the equilibrium 
vortex state is not entangled below the
irreversibility line of the high-$T_c$ superconductors or in a large part of 
the vortex-liquid phase above this line.
 Also, the solution for an infinite straight
screw dislocation is found, and used to give a limiting form for the free
energy cost of very large braids. This depends on the area 
enclosed by the braid as well
as its perimeter length. 

\end{abstract}

\section{Introduction}

There has been a lot of interest in the effects of thermal fluctuations on the
behaviour of flux-lines ever since the discovery of the high-$T_c$
superconductors. The high temperatures and fields at which these new materials
remain in the mixed state, and their small coherence lengths, mean that these
fluctuations may be strong enough to totally alter the nature of the vortex
state from the mean-field solution of an Abrikosov lattice. Debate and
controversy has continued on the possibilities of different vortex phases above
and below the irreversibility line (at which the vortices become depinned). We
believe that the results of the calculations in this paper shed some light on
certain questions on the vortex state. Our primary conclusion is that there is 
very little entanglement near and below the irreversibility line.

In this work, we start with the triangular vortex-lattice, and consider the free
energy costs of topological defects within the lattice. Because of thermal
fluctuations, the spontaneous formation of defects of finite free energy cost
is important in the description of the vortex state. These defects may 
include `crossings' where two or more lines may pass through the same point, and
`braids' where a collection of lines twist around each other as one moves along
the field direction. Our approach
is always to begin with the Abrikosov lattice of vortices that corresponds
to the  mean-field  solution for the Ginzburg-Landau
free energy functional within the lowest Landau level 
approximation\cite{Eilenberger,Fetter}. We
realize of course that the LLL approximation only holds strictly in the high
field regime of the superconducting phase diagram, just below $H_{c2}$, and
even 
that the existence of this lattice may be questioned. Having set up the perfect
lattice of vortices, we than allow a restricted number of the vortex-lines to
move subject to certain boundary conditions that define a topological
defect, and then find the minimum free energy cost of such a defect.

The free energy cost of a crossing configuration provides
an estimate of the energy barrier for vortex lines to cross through each other,
or to cut and reconnect. Accurate estimates of the crossing energy barrier in
the different regions of the $H$-$T$ phase diagram are desired because of
their relevance to the vortex dynamics of high-$T_c$ superconductors. 
The energy barrier, $U_{\times}$, to crossing/cutting of vortex lines is a 
fundamental parameter of the entangled state \cite{Cates,Marchetti}. 
 This has encouraged calculations of this energy in different
approximations \cite{Wilkin,Carraro}.                           
The calculation in this paper is the first to consider the effects on 
$U_{\times}$ of the surrounding triangular lattice. 

The braid defects are examples of the sort of configurations one expects to
find in the entangled vortex state. As well as affecting the response to
pinning centers, they will have interactions with any component of current
parallel to the applied field. The growth of braid defects may be a source of
dissipation of such longitudinal currents.
An important result of our calculation is that there are no stable two-line or
three-line braid defects, at least within our approximations. The smallest
braid defect that is stable, but which has a relatively high free energy cost,
is the six-line braid. The non-existence of stable
braid configurations with small free energies has the consequence that in
a sample of typical size of high-$T_c$ superconductor there is a large region
of the \hbox{$H$-$T$ phase} diagram at which there will be no braid defects present in
thermal equilibrium. We estimate that there will be no braids for values of the
reduced temperature far above the irreversibility line and quite close to the
$H_{c2}$ line (see Section~\ref{sec:noent}). If there are no braid
defects present in a given sample then the vortex system is {\em disentangled}.
This is contrary to the picture of an entangled vortex liquid suggested by 
Nelson \cite{Nelson}. 

In Section~\ref{sec:method} we describe the formulation of our calculations.
Section~\ref{sec:small} contains our calculations and results for crossing and
braid defects
involving two, three, six and twelve lines. In
Section~\ref{sec:large} we extend our method to large scale defects of the
Abrikosov lattice. The free energy cost of an infinite straight screw
dislocation is calculated, and the result is used to find a limiting form for
the free energy cost of very large braids. This free energy is found to depend 
on the area
enclosed by the braid as well as the perimeter length. 
In Section~\ref{sec:appl} it is shown how our results for small braids
lead to the conclusion that entanglements are not important in a large region
above and below the irreversibility line of high-$T_c$ superconductors of
typical sizes. Other applications of the calculations are also suggested
in Section~\ref{sec:appl}.

\section{Method of the Calculations}  \label{sec:method}

For the sake of clarity, and to set the notation used, the general method
behind the calculations is now described. First, the Ginzburg-Landau theory is
formulated and the Lowest Landau Level is defined. Then the ground state
corresponding to an Abrikosov triangular lattice is presented, and it is shown
how to construct deviations from the perfect lattice which remain in the LLL.
The thermodynamics of flux-lines within a type-II superconductor are well
described for fields $H$ near $H_{c2}(T)$ by Ginzburg-Landau theory. Within
this theory, the free energy density is expanded in terms of the
superconducting order parameter and its derivatives, 
\begin{equation} 
{\cal F}\{\psi\}=\sum_{\mu=1}^3\frac{1}{2m_{\mu}} 
{|(-i\hbar\partial_\mu
 -2e\hbox{ A}_\mu)\psi|}^2 +\alpha
 {|\psi|}^2
+ \frac{\beta}{2}{|\psi|}^4 + \frac{b^2}{2\mu_0} +const,\\
\end{equation}  
 where $\hbox{\bf b}=\hbox{\boldmath $\nabla$} \times \hbox{\bf A}$
 is the microscopic flux density. This equation applies for general anisotropy
but only applies to a homogenous system (layering effects are ignored). 
For relevance to layered superconductors, we consider a limited anisotropy with
$m_x=m_y=m_{ab}$ and $m_z=m_c$. For a uniform external field 
$\hbox{\bf H}_0$ parallel to the z-axis, it is the Gibbs free energy,
${\cal G}\{\psi\} =F_0+{\cal F}\{\psi\} - \hbox{\bf b.}\hbox{\bf H}_0$ that
controls the properties of the system \cite{Ruggeri}.       

  Our main approximation (valid for high fields) is to restrict the order
parameter to the `Lowest Landau Level' (LLL) subspace, defined by
\begin{equation} 
\label{lllcondition}
    \Pi \psi =0,
\end{equation}   
where $\Pi =\Pi_x +i\Pi_y$, $\Pi_x=-i\hbar 
(\partial/\partial x)-2eA_x$, and
$\Pi_y=-i\hbar (\partial/\partial y) -2eA_y$.
Within this restriction, substitution into the Gibbs free energy leads to an
effective free energy density of \cite{Ruggeri}
\begin{equation}
\label{free} 
 {\cal F}\{\psi\}= \alpha_H {|\psi|}^2   +\frac{\beta_K}{2} {|\psi|}^4+    
 \frac{\hbar^2}{2m_c} {\left|\frac{\partial\psi}{\partial z}\right|}^2,
\end{equation}
 where $\alpha_H =\alpha + (e\hbar/m_{ab})\mu_0 H_0$, and
$\beta_K=\beta-\mu_0 (e\hbar/m_{ab})^2$.
Here, $\alpha_H$ is our reduced temperature variable, which is negative below
the $H_{c2}$ line and positive above. It is useful to make the transformations
\begin{eqnarray}
\label{transform}
\psi\rightarrow \tilde{\psi}&=& {\left(\frac{\beta_K}{|\alpha_H|}\right)}
^{\frac{1}{2}}\psi\nonumber\\
z\rightarrow h&=& {\left(\frac{2m_c|\alpha_H|}{\hbar^2}\right)}^{\frac{1}{2}}z.
\end{eqnarray}
Thus, $h$ is the distance along the $c$-axis in units of the mean field 
correlation length in this direction, $\xi_c=\sqrt{\hbar^2/2m_c|\alpha_H|}$.
Substituting (\ref{transform}) into (\ref{free}) and integrating over all space
(for an infinite bulk superconductor) gives a total free energy
\begin{equation}
\label{totalfree}
F=  {\left(\frac{\hbar^2}{2m_c}\right)}^{\frac{1}{2}} \frac{{|\alpha_H|}^
{\frac{3}{2}}}{\beta_K}
\int dh \int d^2r \left\{ -{|\tilde{\psi}|}^2 +\frac{1}{2}
 {|\tilde{\psi}|}^4
+ {\left|\frac{\partial\tilde{\psi}}{\partial h}\right|}^2
\right\}.
\end{equation}         
   
A consequence of using the 
LLL approximation is that solutions of (\ref{lllcondition}) have the general
form in the plane perpendicular to the magnetic field \cite{Eilenberger}
\begin{equation}
\psi_0 (x,y)=f(z)e^{-{\pi y^2}/{\eta}},
\end{equation}
where $z$ is not the third dimension in space, but the complex variable, 
$z=x+iy$, and  
$\eta=\Phi_0/B$, ($B\equiv \langle b\rangle$ is the mean magnetic flux
density, and $\Phi_0=h/2e$ is the flux quantum).
  $f(z)$ is an analytic function of $z$. Any
function of the above form is a function within the LLL subspace. $f(z)$ may be
written quite generally in a product form
\begin{equation}
\label{product}
f(z)\propto \prod_i(z-z_i),
\end{equation}
where the zeros in $f(z)$, $z=z_i$, correspond to the positions of vortices
within the superconductor. Of course it is well known that the order parameter
that minimizes the free energy (\ref{free}) in the mixed state 
is a triangular periodic lattice
of flux-lines \cite{Fetter}. The corresponding function $f(z)$
for such a lattice is a Jacobi theta function \cite{Eilenberger}
\begin{equation}
\label{jacobi}
\psi_0 (x,y)=\phi(\hbox{\bf r}|\hbox{\bf 0})=
Ce^{-\frac{\pi y^2}{\eta}}\vartheta_3\left(\frac{\pi z}{l},
\frac{\pi\tau}{l}\right).            
\end{equation}
Properties of $\phi(\hbox{\bf r}|\hbox{\bf 0})$ 
and related functions $\phi(\hbox{\bf r}|\hbox{\bf r}_0)$ that 
span the LLL subspace are given
in \cite{Eilenberger}. The function ${\left|\phi(\hbox{\bf r}|\hbox{\bf 0})
\right|}^2$
 has the periodicity $\hbox{\bf r}_I$ and $\hbox{\bf r}_{II}$ where
$\hbox{\bf r}_I=l(1,0)$,
$\hbox{\bf r}_{II}=l(1/2,\sqrt{3}/2)$, and
$\tau=x_{II} +iy_{II}={l}/{2}+il\sqrt{3}/2$.
$l$ is the distance between neighboring zeros (vortices) in 
$\phi(\hbox{\bf r}|\hbox{\bf r}_0)$.
The area of the unit cell must contain one quantum of flux, so that
$\eta=l^2\sqrt{3}/2$.
The prefactor in (\ref{jacobi}) is found to be $C=3^{1/8}$, if we use  
the normalization condition \cite{Eilenberger}
\begin{equation}
\label{normal}
\langle {|\psi_0|}^2\rangle =\int_{\mbox{\scriptsize \em unit cell}}
\frac{d^2r}{\eta} {\left| \phi(\hbox{\bf r}|\hbox{\bf 0})
\right|}^2=1. 
\end{equation}
To find the free energy of the Abrikosov lattice, we substitute
$\tilde{\psi}(x,y,h)=K\psi_0(x,y)$ 
into (\ref{totalfree}). As ${\left|\psi_0\right|}^2$ is periodic, 
we can integrate over one unit cell to find the average free energy density
\begin{equation}
f=  {\left(\frac{\hbar^2}{2m_c}\right)}^{\frac{1}{2}} \frac{{|\alpha_H|}^
{\frac{3}{2}}}{\beta_K}
\int_{\mbox{\scriptsize \em unit cell}} \frac{d^2r}{\eta} \left\{
 -K^2{\left|\psi_0\right|}^2  +K^4{\left|\psi_0\right|}^4
\right\}.
\end{equation}   
This is minimized with respect to $K$ (using (\ref{normal})) by the condition
$K^2=1/\beta_A$, where $\beta_A$ is the Abrikosov parameter
$\beta_A=\langle {\left|{\psi_0}\right|}^4\rangle /
{\langle {\left|{\psi_0}\right|}^2\rangle}^2 \simeq 1\cdot 1596$.
The free energy per flux-line per unit length for the Abrikosov lattice is
therefore given by
$f_L=-{{\alpha_H}^2}/{2\beta_K\beta_A}$.

As any LLL function has the general form (\ref{product}), including $f(z)=
\vartheta_3\left(\pi z/l,\pi\tau/l\right)$ (see 
Appendix~\ref{ap:theta}), we can therefore construct a function 
representing the order
parameter for a triangular lattice with one flux line displaced from $z=z_0$
 to $z=\zeta_0$, by
\begin{equation}
\label{displaced}
\psi_{\mbox{\scriptsize \em displaced}}(\hbox{\bf r})=\psi_0(\hbox{\bf r})
\frac{(z-\zeta_0)}{(z-z_0)}.
\end{equation} 
This cancels out the first order zero in $\psi_0$ at $z=z_0$ and 
replaces it with a first order zero at $z=\zeta_0$. Because this 
new function has the form of (\ref{product}), it
is still within the LLL subspace, even though it is obviously not the
configuration of lowest free energy (which is $\psi_0$). (A similar method to
this was used by Brandt\cite{Brandt}).
(\ref{displaced}) can be generalized to any number of displaced lines, to form
order parameters for various defects of the perfect lattice within the LLL
subspace. The free energy cost of any such defect is given by integrating 
over all space the
difference in the free energy densities of the defect and the ground state:
\begin{equation} \label{defect}
\Delta F_{\mbox{\scriptsize \em defect}}=  
{\left(\frac{\hbar^2}{2m_c}\right)}^{\frac{1}{2}} \frac{{|\alpha_H|}^
{\frac{3}{2}}}{\beta_A\beta_K}
\int dh \int d^2r \left\{ \cal{F}\left(\psi_{\mbox{\scriptsize \em defect}}
\right) - \cal{F}\left(\psi_0\right) \right\},
\end{equation}
where   
\begin{equation}
\label{freeint}
\cal{F}(\psi)= -{|{\psi}|}^2 +\frac{1}{2\beta_A}
 {|{\psi}|}^4
+{\left|\frac{\partial{\psi}}{\partial h}\right|}^2.
\end{equation}
      
In all the problems described in this paper, the procedure is to define a
defect in the perfect lattice by allowing the positions of some of the vortices
to vary with $h$ subject to certain boundary conditions leaving the rest of the
vortices fixed in their positions in the triangular lattice. The total
free energy of the defect is then minimized with respect to the positions of
the chosen vortices, subject to these boundary conditions. This is done by
expanding the free energy density as a polynomial in a set of variables that
describe the coordinates in the $x$-$y$ plane of the chosen lines, say
$\{\zeta_i(h)\}$ for the vortices labelled by $i$. The coefficients in the expansion
will be functions of $x$ and $y$ only (as there is no $h$ dependence of
$\psi_0$). These coefficients must be integrated over the $x$-$y$ plane to
give the free energy per unit length along $h$, that depends on
 $\{\zeta_i(h)\}$ and
their first derivatives. The form of the variables $\{\zeta_i(h)\}$ 
that minimizes
the total free energy will be solutions to Euler-Lagrange non-linear
differential equations. By solving these the
lowest free energy cost of a given type of defect can be found.

One of our main approximations is that we do not allow the lines not directly
involved in the defect to move in response to the presence of the defect (i.e.\
the relaxation of the surrounding lattice). This may seem like a poor
approximation that will seriously over estimate the free energies of the real
defects. However, we have found that this is not the case when we allow the
nearest neighboring lines to move (as in Section~\ref{sec:relax}) and we 
believe
that relaxation of the lattice will only slightly reduce our values of defect 
free energy costs. The reason for this is that we are considering a lattice of
vortex {\em lines}, which cost energy to tilt with respect to the field
direction, so if we consider a localized defect taking place over a small
length scale, $L$, then the surrounding lines would have to tilt considerably
over the distance $L$ if they are to reduce their interaction energy with the
defect lines, and this will cost too much tilt energy.

An important general result of our method is that for any defect of the
ground state made from changing the positions of $n$ of the vortices, the free
energy cost will diverge logarithmically with the total size of the system,
unless the $n$ lines {\em move symmetrically} about their 
`mean midpoint'\cite{Brandt}. That
is, if $n$ lines move from their ground state values $\{z_i\}$ to the new positions
$\{\zeta_i(h)\}$, then the condition for a non-divergent free energy is:
\begin{equation}  \label{symcond}
\sum_{i=1}^n \zeta_i(h) = \sum_{i=1}^n z_i,
\end{equation}
i.e.\ if we define the `center' of the defect as the vector sum of the coordinates
of the ground state lattice position, $z_i$, then this condition says that the
vectors of the coordinates of the defect line must always sum to this `center'.
An important point to note is the restriction in this derivation to a defect of
$n$ lines. If one wanted to consider a defect that by definition could not
satisfy the symmetry relation (\ref{symcond}) then a finite free energy could still be
obtained, but only by allowing the `relaxation' of other lines in the system.
The motion of these extra lines would be so as to ensure the symmetry condition
for the total number of moving lines, and through this cancel out the
logarithmic divergence. However, if it is possible for a simple $n$-line defect
to satisfy (\ref{symcond}), then the free energy cost will depend only on the
{\em locally} surrounding lattice.
We take advantage of this result in
the following calculations to reduce the number of free parameters used for
describing a given defect.

\section{Calculations for Small Defects} \label{sec:small}
\subsection{Two Lines Crossing}

\begin{figure}[htbp]           
\epsfxsize=15cm
\begin{center}
\leavevmode\epsfbox{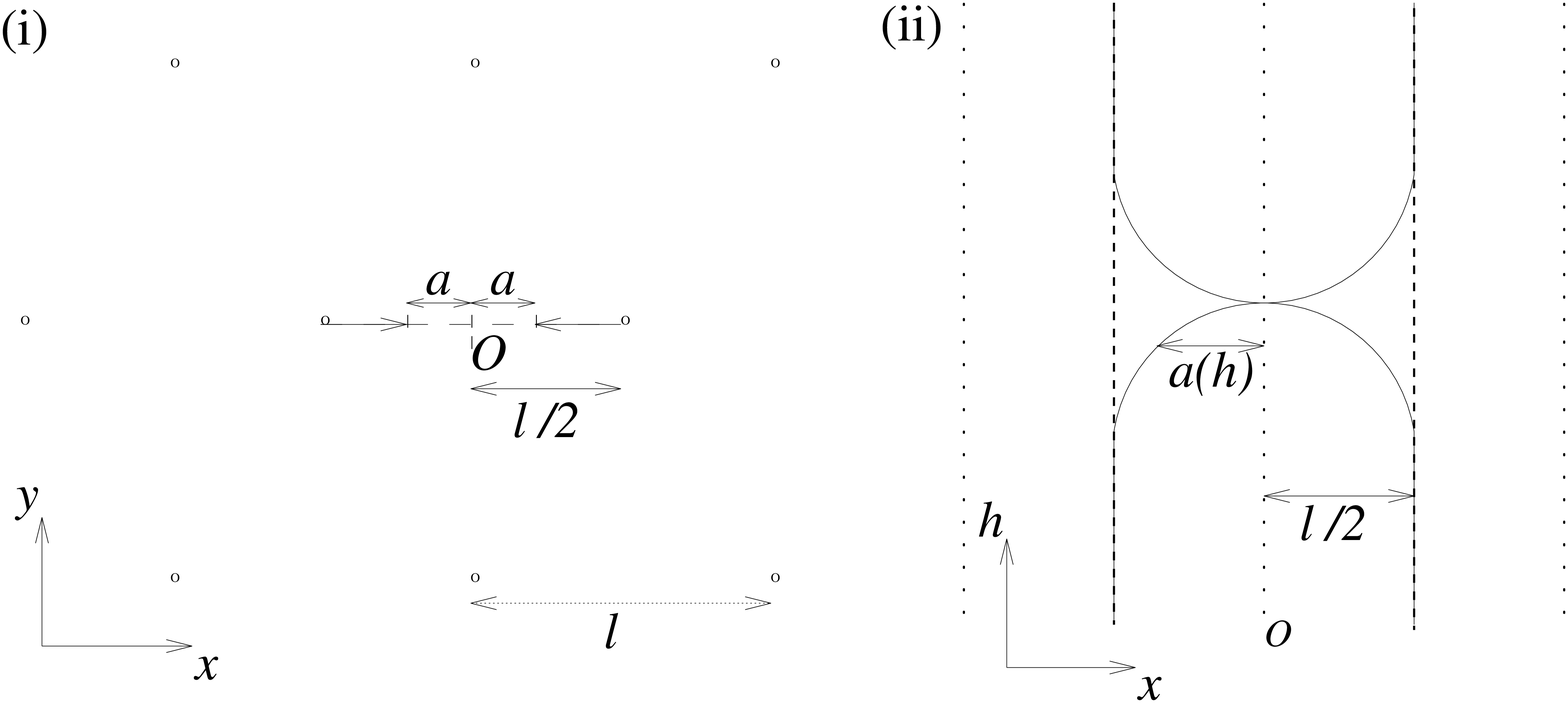}
\\
\caption{ 
(i) Cross-section of the vortex lattice on the $x$-$y$ plane, showing the path
along which two neighboring lines may cross. The ground state positions of
vortices are marked by the small circles. (ii) Schematic side-view of the vortex
lattice with two lines crossing. The solid lines are in the plane of the paper
and the dotted lines represent vortices just behind/in front of this plane.
\label{fig:1}}
\end{center}
\end{figure}

The first defect to which the above method is applied is the case of two
neighboring lines within the Abrikosov lattice moving together to meet at a
point. The importance of this defect is that it is the
configuration providing an energy barrier to the `cutting' or `reconnecting' of
two vortices. The simplest way to estimate the energy of two lines crossing is
to keep all the surrounding lines in their original positions in the lattice,
and to assume that the two lines move symmetrically towards each other (see
Fig~\ref{fig:1}). This avoids a divergence in the free energy cost, and
results in an order parameter depending on one parameter only.

\begin{figure}[htbp]           
\epsfxsize=8cm
\begin{center}
\leavevmode\epsfbox{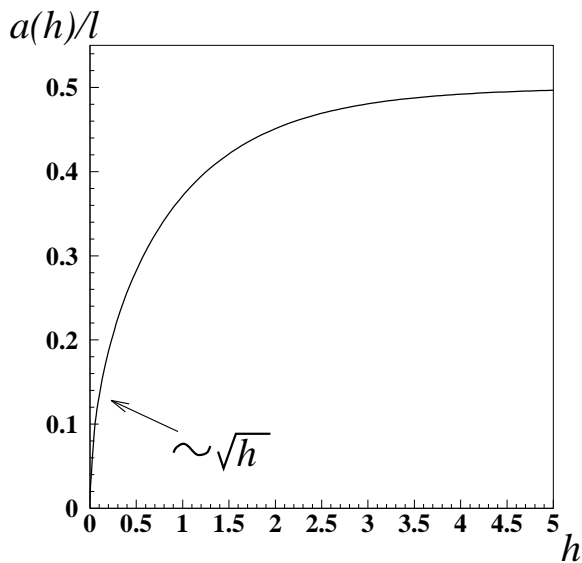}
\caption{
The form of $a(h)$ that minimizes the free energy subject to the 
crossing boundary condition that $a(0)=0$. The expected form  
as $h\rightarrow 0$ is shown.\label{gr:2} }
\end{center}
\end{figure}

The calculation for this simple defect is described in more detail than the
other problems in order to demonstrate the general procedure of these
calculations.
If we take the origin, $O$, of the complex $z$-plane to be at the midpoint of the two
lines, then we have two vortices (= zeros in the
order parameter) which move from $z=\pm {l}/{2}$ to $z=\pm a(h)$, 
with the boundary conditions $a(\pm\infty)=\pm{l}/{2}$.
Using (\ref{displaced}), the LLL order parameter for two lines crossing 
may be written
\begin{equation}
\label{abriko}
\psi_{2}(x,y,h)=\psi_0(x,y)\frac{(z+a(h))(z-a(h))}{(z+\frac{l}{2})
(z-\frac{l}{2})}.
\end{equation} 
The subscript, $2$, indicates the number of lines allowed to deviate from their
positions in the ground state triangular lattice.
It is now necessary to find an expression for the free energy change of the
displaced lines compared to the undistorted lattice as a functional of $a(h)$,
then minimize this functional to find the correct shape of the crossing lines
(i.e.\ the configuration of lowest free energy). Looking at (\ref{defect})
and (\ref{freeint}) one sees that we need to expand the differences 
${|\psi_{2}|}^2 -{|\psi_0|}^2$, and
${|\psi_{2}|}^4 -{|\psi_0|}^4$, as well as 
${\left|{\partial\psi_2}/{\partial h}\right|}^2 $ in terms of $a$. The
coefficients of these expressions   are then integrated over the $x$-$y$ plane
which leads to a free energy per unit length along the $h$-direction of
the form:  
\begin{eqnarray}
\label{poly}
f_2\left\{ a(h)\right\} &\equiv& 
\int \frac{d^2r}{l^2} \left\{ \cal{F}(\psi_2) - \cal{F}(\psi_{0})  
\right\}\nonumber\\   
&=&\sum_{i=0}^4{c^{(2)}_ia^{2i}} +c^{(2)}a^2{\left(\frac{da}{dh}\right)}^2, 
\end{eqnarray}
with the coefficients $c^{(2)}_i$ equal to the $c^{(2)}_{i0}$
given in Table~\ref{tab:c2}, and $c^{(2)}\simeq
94.20/l^4$. 
For example, the coefficient
$c^{(2)}_1$ was found by calculating:
\begin{eqnarray}
c^{(2)}_1&=&2\int \frac{d^2r}{l^2}\, \frac{{|\psi_0|}^2 Re\left(
z^2\right)}{{|z^2-(\frac{l}{2})^2|}^2}\:
-\frac{2}{\beta_A}\int \frac{d^2r}{l^2}\, \frac{{|\psi_0|}^4 Re\left(
z^2\right){|z|}^2}{{|z^2-(\frac{l}{2})^2|}^4}\\
&\simeq& -5\cdot 201 /l^2. 
\end{eqnarray}
Note that all of the integrals that make up these coefficients must be 
convergent, as the denominator of the
integrands always grows (for large $z$) as at least two powers of $|z|$ higher 
than for the numerator (there is no logarithmically divergent coefficient as
the symmetry condition (\ref{symcond}) is satisfied).
\begin{figure}[tbp]           
\epsfxsize=8cm
\begin{center}
\leavevmode\epsfbox{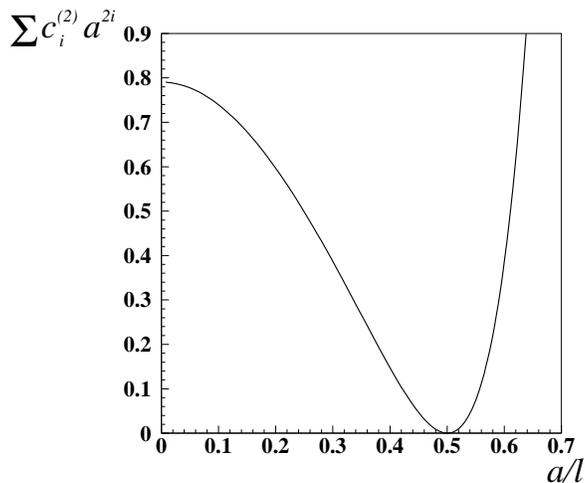}
\caption{ The potential term, $\sum_i c_i^{(2)}a^{2i}$, for two lines crossing.
\label{gr:1}}
\end{center}
\end{figure}
The first term in (\ref{poly}) represents the potential energy cost per unit
length with the two lines displaced but still straight. The form of this term
is shown in Fig~\ref{gr:1}. The second term is the tilt energy of the two lines. It
increases with the distance $a$, which is consistent with an attractive force
between the anti-parallel components of the vortex segments in the two lines.
One could think of the equation (\ref{poly}) in analogy with a Lagrangian
density in classical mechanics, of the general form $\cal{L}=T-V$ for a body
moving in one dimension, $a$, and with `time'
equivalent to the $h$-direction.
The total free energy is proportional to the integral over all $h$ of
(\ref{poly}). Applying the Euler-Lagrange equation for stationary values of a
functional, and integrating once, one arrives at the equation for the
form of $a(h)$ that minimizes the free energy change,
$f_2-a'({\partial f_2}/{\partial a'})=\mbox{\em const}$.
Substituting (\ref{poly}) and applying the boundary conditions gives
\begin{equation}
c^{(2)}a^2{\left(\frac{da}{dh}\right)}^2= \sum_{i=0}^4{c^{(2)}_ia^{2i}}.
\end{equation}
This can now be integrated up from the point of crossing, $a=0$ and (say) $h=0$,
to give the correct form of $a(h)$. As  $h\rightarrow 0$ this will tend to 
$a(h)\simeq (4c_0^{(2)}/c^{(2)})^{1/4} \sqrt{h}$ (this form at small $a$ has
been found before\cite{Wilkin}).
The result is shown in Fig~\ref{gr:2}. The
stationary form of $a(h)$ can now be substituted into (\ref{poly}) and
(\ref{defect}), and the
integral over $h$ performed to find the total free energy change for crossing.

We find 
$\int dh\; f\!\left\{ a(h)\right\}\simeq 2\cdot 32$, which gives
\begin{equation}
\label{change}
\Delta F_{2\times}
= {\left(\frac{\hbar^2}{2m_c}\right)}^{\frac{1}{2}} \frac{{|\alpha_H|}^
{\frac{3}{2}}}{\beta_K\beta_A }
\frac{2\Phi_0}{\sqrt{3}B}
\times 2\cdot 32.
\end{equation}
To put this in a simpler form, we use the dimensionless factor
$\alpha_T$ defined by
\begin{equation}
\label{alphat}
\alpha_H={\left( \frac{\beta_Ke\mu_0Hk_BT\sqrt{2m_c}}{4\pi\hbar^2}\right) }
^{\frac{2}{3}}\alpha_T.
\end{equation}
Substituting (\ref{alphat}) into (\ref{change}) gives
\begin{equation}  \label{result}
\Delta F_{2\times}=0\cdot 58\;k_BT{|\alpha_T|}^{\frac{3}{2}}.  
\end{equation}   
All of the following free energies will also be quoted in this form. However,
there is at present a variety of different units used in the literature. For
comparison, we can also write $\alpha_T$ in terms of the temperature, the
field, and
the Ginzburg number $Gi$. (The Ginzburg number is a useful parameter that
describes the strength of thermal fluctuations of the superconducting order
parameter \cite{Blatter}). 
 We use the definition of $Gi$ given by \cite{Blatter}:
\begin{equation}
Gi=\frac{1}{2{(8\pi)}^2}{\left( 
\frac{2k_BT_c} {\mu_0{H_c}^2(0)\xi_{ab}^2(0)\xi_c(0)}\right)}^2.
\end{equation}
The factor $(1/8\pi)^2$ enters as we change the units of magnetic energy from
c.g.s.\ to S.I. Using standard relations for the
coherence length and critical fields  we find that:
\begin{equation}
{|\alpha_T|}^\frac{3}{2}=\frac{\sqrt{2}{(1-t+\hbox{h})}^\frac{3}{2}}
{{Gi}^{1/2}\hbox{h}t}.
\end{equation}
$t$ and $\hbox{h}$ are the dimensionless temperature and field $t=T/T_c$ and
$\hbox{h}=H/H_{c2}(0)$ where $H_{c2}(0)$ denotes the straight line 
extrapolation to zero temperature of the $H_{c2}$ line when the field is
applied along the $c$-axis.

\subsection{Two Lines Crossing with Relaxation of Nearest Lines} 
\label{sec:relax}

\begin{figure}[tbp]           
\epsfxsize=10cm
\begin{center}
\leavevmode\epsfbox{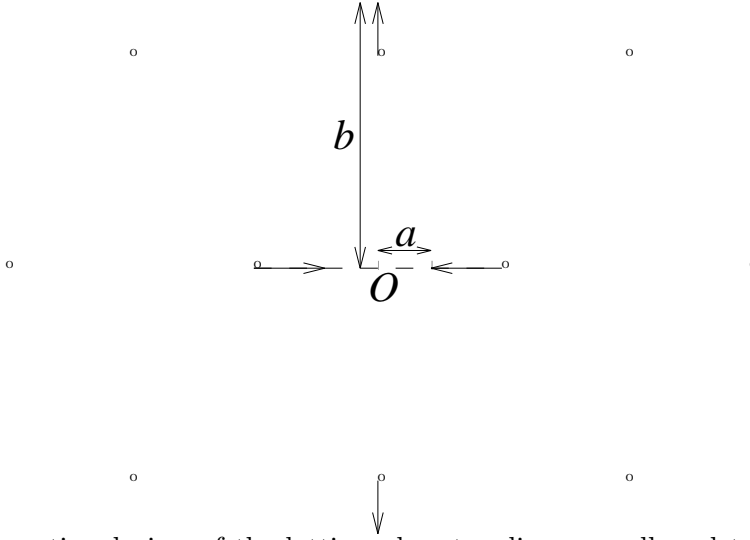}
\caption{  
Cross-sectional view of the lattice when two lines are allowed to cross, and
the two nearest neighbors move in response.
\label{fig:2}}
\end{center}
\end{figure}

\begin{figure}[tbp]           
\epsfxsize=8cm
\begin{center}
\leavevmode\epsfbox{    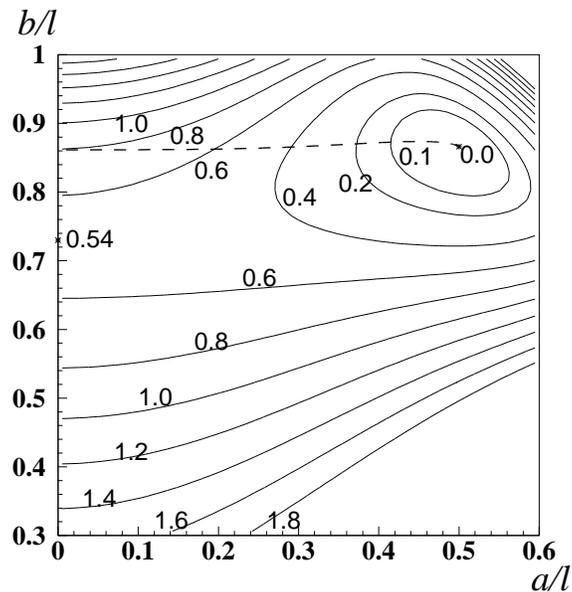}
\caption{ The potential term, $C_4(a,b)=\sum_{i,j} c_{ij}^{(4)}a^{2i}b^{2j}$, 
for two lines crossing with the two nearest other lines also allowed to move.
The dashed line shows the actual path taken in the full solution.
 \label{gr:3}}
\end{center}
\end{figure}

The result in (\ref{result}) is only an upper bound on the
actual energy for the two lines to cross. This is because we have ignored any
relaxation that the remaining lines in the lattice may undergo in reaction to
the crossing so as to decrease
the free energy cost of the defect. In order to test how good an
estimate this result is, we now investigate the extent that 
 allowing the surrounding vortices to move may change the free energy. 
It seems likely that it
is the lines closest to the crossing region which will move the most in
this defect, and therefore affect the energy of the crossing the most. As a
second approximation to (\ref{result}), a calculation was performed where the
two lines closest to the crossing center, above and below, were allowed to move
symmetrically in response to the crossing (see Fig~\ref{fig:2}).
The new order parameter with this allowed motion of the four lines is:
\begin{equation}
\psi_{4}(x,y,h)=\psi_0(x,y)\frac{(z^2-{a(h)}^2)(z^2+{b(h)}^2)}
{(z^2-{(\frac{l}{2})}^2)
(z^2+{(\frac{\sqrt{3}l}{2})}^2)}.
\end{equation} 
The boundary conditions of this problem are that $a(\pm \infty )=l/2$, 
$b(\pm \infty )=l\sqrt{3}/2$, and $a(0)=0$. Following the
general procedure of the first calculation, the free energy per unit length as
a function of $a(h)$ and $b(h)$ was found to be:
\begin{eqnarray}  \label{freeex}
f_4\{ a(h),b(h)\} &=&
\sum_{i,j=0}^4 c^{(4)}_{ij}a^{2i}b^{2j} 
+a^2{\left( \frac{da}{dh}\right) }^2\sum_{i=0}^2u_ib^{2i}\nonumber\\ 
&&+b^2{\left( \frac{db}{dh}\right) }^2\sum_{i=0}^2v_ia^{2i}  
+ab\frac{da}{dh}\frac{db}{dh}\sum_{i,j=0}^1w_{ij}a^{2i}b^{2j}.
\end{eqnarray}      
  The coefficients $c^{(4)}_{ij}$, $u_i$, $v_i$, and $w_{ij}$ are given in
Table~\ref{tab:c4}.
Again the first term can be viewed as a potential term due to the interaction
between straight lines. Its form in the $a$-$b$ plane is shown in 
Fig~\ref{gr:3}.
 The other terms  describe the tilt energies of the 
lines, with a fairly complex dependence on the positions of the lines.
Looking at the form of the potential term 
$C_4(a,b)=\sum_{i,j=0}^4 c^{(4)}_{ij}a^{2i}b^{2j}$ in Fig~\ref{gr:3}
we find that 
the interaction of straight lines in the LLL Abrikosov lattice is not at all as
we might expect. One would think that parallel vortices always have a repulsive
force between each other, but instead Fig~\ref{gr:3} 
shows the surprising feature that
if the two crossing lines are placed at the midpoint, the two nearest
neighboring lines are actually pulled towards the midpoint! (i.e.\ the free
energy is reduced by moving towards the center.) It is impossible to explain
this effect by any simple two-body interaction between vortex-lines.
The minimum value of $C_4(a,b)$ when $a=0$ is  $C_4(0,0.73)\simeq 0.54$ which
is considerably less than the value when the two extra lines remain at their
original positions, $C_4(0,\sqrt{3}/2)\simeq 0.79$. Therefore, allowing
nearby lines to move will make a large difference to the potential terms for
two lines crossing. However, one must also take into account the bending terms
of all the lines involved in the full solution. 
Now, to find the configuration of $a(h)$ and $b(h)$ that minimizes the free
energy cost of two lines crossing, we have to solve two coupled Euler-Lagrange
equations of the form:
\begin{equation}\label{euler}
\frac{\partial f_4}{\partial x_i}-\frac{d}{dh}\left( \frac{\partial f_4}
{\partial x_i'} \right) =0.
\end{equation} 
Substituting (\ref{freeex}) into (\ref{euler}) leads to two coupled equations
that  were solved numerically with a relaxation method, using a software 
package. The resulting configuration from the solution to (\ref{euler})
 is only slightly different from the configuration when the
nearby lines are kept fixed. The two nearest lines move in by $\sim 0.03\, l$ 
at $h=0$ (the actual path is shown in Fig~\ref{gr:3}). 
The difference in energy is very small, with the same answer
to the accuracy quoted as (\ref{result}).
As we had expected the movement of the nearest lines to those crossing to
affect the free energy change the most, it seems that the initial approximation
of not allowing any of the surrounding lines to move is quite good, and we are
fairly confident that the result (\ref{result}) is close to the exact
answer. The reason that the relaxation of surrounding lines has so little
effect on the crossing energy is that it costs too much free energy to bend
these extra lines over the short distance along $h$ that crossing takes place,
so that the minimum in the potential term is never reached. This is a
consequence of having a lattice of lines rather than pancake vortices.

\subsection{Search for a Two Line Braid}

The other extension for the two-line problem is  not to restrict the lines to
moving in  one direction only but to allow for the possibility that the two 
lines may `braid' around each other (by braid it is meant that the pair of 
lines twist around each other by half a rotation as we move along the
 direction of the lines). One of the original motivations to calculating
the crossing energy was that it was thought to be an energy barrier to two
topologically distinct configurations. For example, if one considers the two
distinct braid configurations consisting of either a twist by $+\pi$ or by 
$-\pi$,
then to go between the two states one has to pass through a configuration with
two lines passing through each other. In fact, we show that there are no
stable two-line braid-defects of the LLL Abrikosov solution (although the
two-line crossing still remains an important energy barrier between larger
entanglements).

\begin{figure}[htbp]           
\epsfxsize=10cm
\begin{center}
\leavevmode\epsfbox{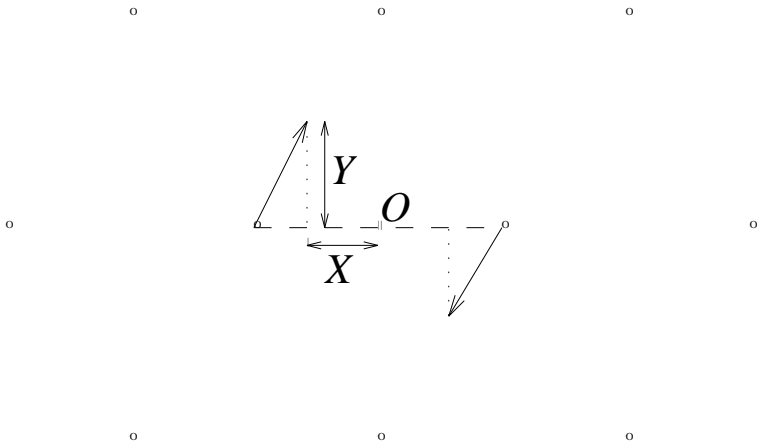}
\caption{
Cross-section of the vortex-lattice with two lines allowed to braid around each
other.
 \label{fig:3}}
\end{center}
\end{figure}

We again allow the two neighboring lines to move symmetrically with respect
to each other, but throughout the $x$-$y$ plane. The positions of the two lines
are given by $\left( X(h),Y(h)\right)$ and $\left( -X(h),-Y(h)\right)$
respectively (see Fig~\ref{fig:3}). The new order parameter is given by
\begin{equation}
\psi_{2}(x,y,h)=\psi_0(x,y)\frac{(z^2-{\left( X+iY\right)}^2)}
{(z^2-{(\frac{l}{2}})^2)}. 
\end{equation} 

\begin{figure}[htbp]           
\epsfxsize=8cm
\begin{center}
\leavevmode\epsfbox{    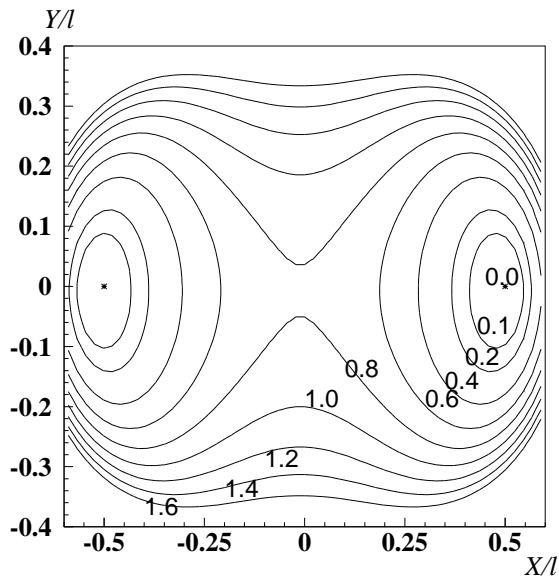}
\caption{ The potential term, $C_2=\sum_{i,j} c_{ij}^{(2)}X^{2i}Y^{2j}$, for
two lines allowed to braid or cross. A braid will correspond to going from the
minimum at $(0.5,0)$ around to the minimum at $(-0.5,0)$. The path with lowest
potential will be along the $X$-axis, which corresponds to a
crossing defect.
 \label{gr:4}}
\end{center}
\end{figure}

The boundary conditions for a braid are that
$X(\infty )=l/2$, $X(-\infty )=-l/2$, $Y(\pm\infty)=0$. With a similar
calculation to the one parameter case, the free energy as a function of $X$ and
$Y$ can be derived:
\begin{equation}  \label{freebr}
f_2\{ X(h),Y(h)\} =\sum_{i,j=0}^4 c^{(2)}_{ij}X^{2i}Y^{2j}
+c^{(2)}\left( X^2+Y^2\right) \left( {\left( \frac{dX}{dh}\right) }^2 +
{\left(\frac{dY}{dh}\right)}^2 \right),
\end{equation}                                                   
where $c^{(2)}_{ij}$ are given in Table~\ref{tab:c2}
 (Setting $Y(h)=0$, $X(h)=a(h)$ reduces
(\ref{freebr}) to (\ref{poly}) ). The first term 
$C_2(X,Y)=\sum_{i,j=0}^4 c^{(2)}_{ij}X^{2i}Y^{2j}$
is the interaction term without bending of the lines, shown in Fig~\ref{gr:4}. 
We could now go on as before in trying to solve the Euler-Lagrange
equations to minimize the integral of (\ref{freebr}) subject to the braid 
boundary conditions. Inspection of Fig~\ref{gr:4}
 reveals that there is no
point in this-- there can be no braid solution for this case! This is because
the form of the potential part of the free energy is always increasing as we
increase $Y$ away from the midpoint (the midpoint is a saddle-point of
$C_2(X,Y)$). As the bending term will increase with $(X^2+Y^2)$ for any given
tilt, it is clear that the free energy cost of any two line braid will increase
continuously as the path of the braid moves away from the midpoint. i.e.\ there
are no stationary configurations for the braid boundary conditions other than
the direct crossing already discussed.

One might find it extremely surprising that two straight vortices placed at the
same point will not repel each other along the $Y$ direction as well as the $X$
direction, even in the presence of the surrounding lattice. This seems to be
another non-intuitive feature of the way vortices interact in the LLL. 
It is a possibility that the absence of a braid solution is an artefact
of our approximation of not allowing the surrounding vortices to move. However,
the result of Section~\ref{sec:relax} suggests that these surrounding lines 
should not have a great effect on the two-line defect. Also, a similar 
calculation to that in Section~\ref{sec:relax} was performed where the two 
nearest neighboring lines were allowed to move in response to the motion of 
the two lines along $Y$. This still gave the
same result that the potential term always increases as $Y$ increases.

\subsection{Three- and Six-Line Defects}

\begin{figure}[htbp]           
\epsfxsize=10cm
\begin{center}
\leavevmode\epsfbox{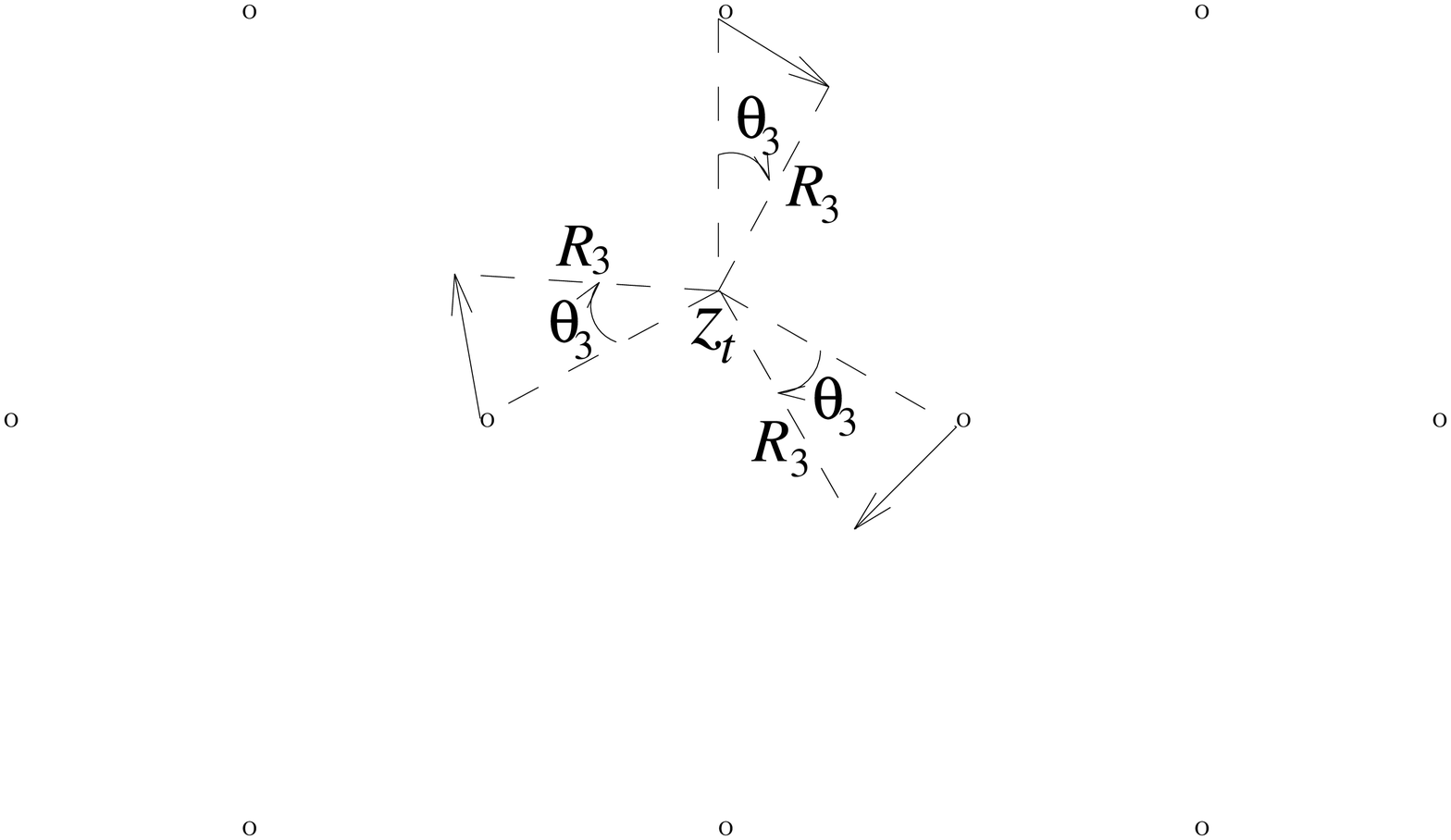}
\vspace{.3in}
\caption{
Cross-section of the vortex-lattice with three lines allowed to braid 
around each other. The center of the defect is labelled $z_t$.       
\label{fig:4}}
\end{center}
\end{figure}

\begin{figure}[htbp]           
\epsfxsize=10cm
\begin{center}
\leavevmode\epsfbox{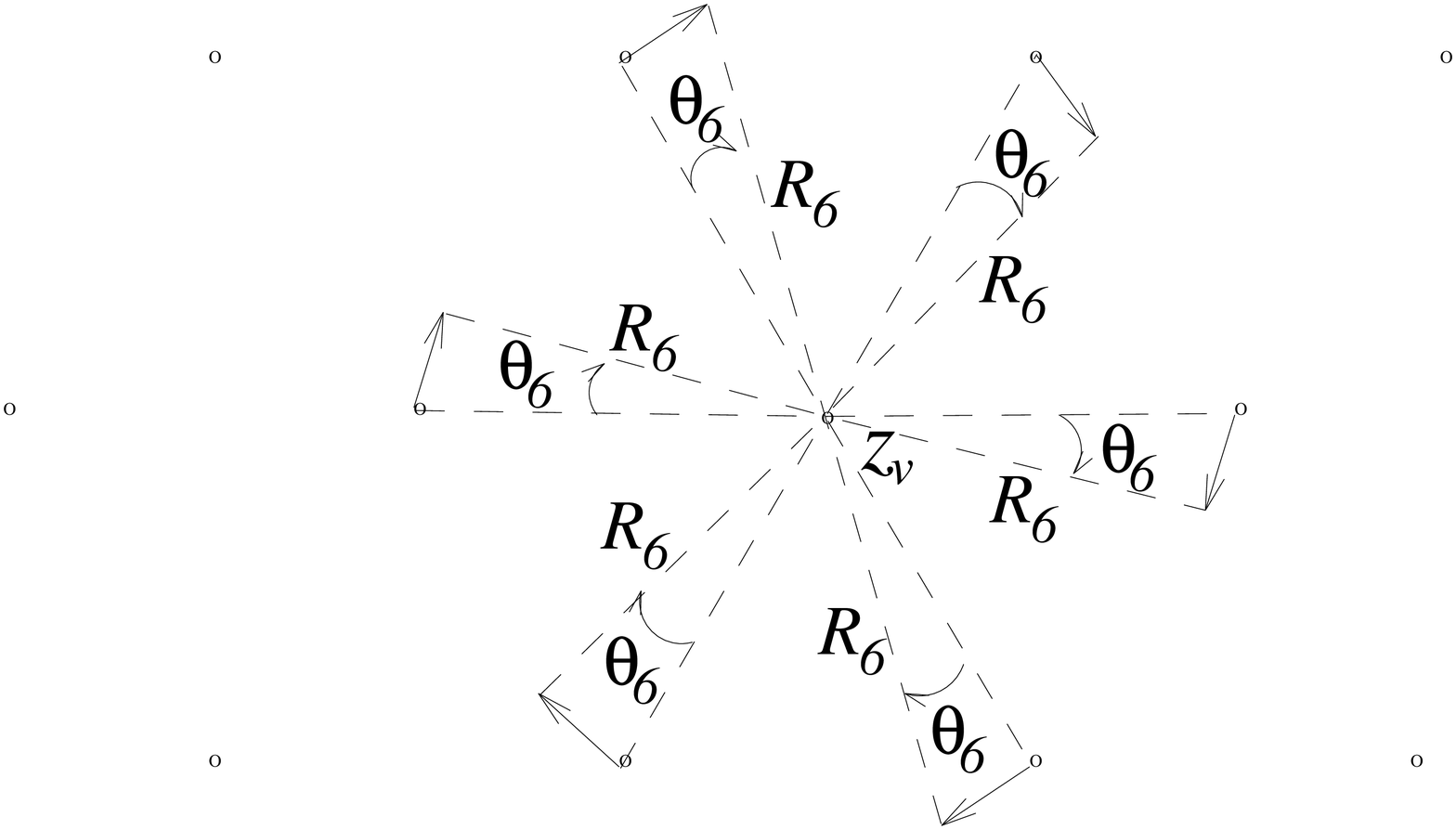}
\caption{
Cross-section of the vortex-lattice with six lines allowed to braid around each
other. The position of the central vortex is $z_v$.   
\label{fig:5}}
\end{center}
\end{figure}

Due to the symmetries of the triangular lattice, there are two other forms of
braid defects which can be handled in the same way as for the two-line
defect. Allowing motion of the three lines that form the smallest triangle (see
Fig~\ref{fig:4}), or the six lines that form a hexagon surrounding one central
line (Fig~\ref{fig:5}). 
If we assume that in the lowest energy configurations the lines in these 
defects move symmetrically about their center, we can write the order parameter
of the defects as:
\begin{eqnarray} 
\psi_{3}(x,y,h)&=&\psi_0(x,y)\frac{({(z-z_t)}^3+i{R_3}^3e^{i3\theta_3})
} 
{({(z-z_t)}^3+i{(\frac{l}{\sqrt{3}}})^3)},\\
\psi_{6}(x,y,h)&=&\psi_0(x,y)\frac{({(z-z_v)}^6-{R_6}^6e^{i6\theta_6})} 
{({(z-z_v)}^6-l^6)},
\end{eqnarray}
 The free energy costs per unit length are of the form:
\begin{equation}     \label{freen}
f_n\{{R_n}(h),\theta_n(h)\} =\sum_{i,j,k} c^{(n)}_{ijk}{R_n}^{3i}
\cos^jn\theta_n \sin^kn\theta_n     
+c^{(n)}\left( {R_n}^{2n-2}{\left( \frac{d{R_n}}{dh}\right) }^2 +
{R_n}^{2n}{\left(\frac{d\theta_n}{dh}\right)}^2 \right),
\end{equation}  
with $n=3,6$. The constants $c^{(3)}_{ijk}$ and $c^{(6)}_{ijk}$
 are given in Table~\ref{tab:c3}. $c^{(3)}\simeq 317.9/l^6$, and 
$c^{(6)}\simeq 117.7/l^{12}$. 
The conditions for a braid are $\theta_n(\infty)=0$, 
$\theta_n(-\infty)=2\pi/n$
and ${R_n}(\pm\infty )$ equal to the ground state value. 
 Again, to minimize the total free energy of given
defects, two coupled E-L equations must be satisfied.
\begin{figure}[tbp]           
\epsfxsize=10cm
\begin{center}
\leavevmode\epsfbox{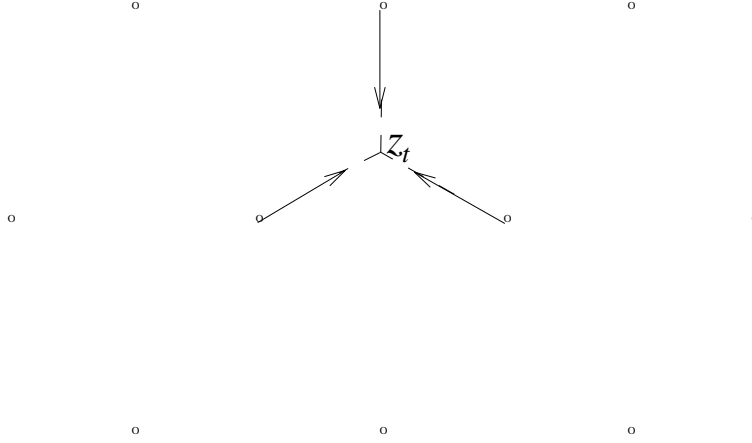}
\vspace{0.5in}
\caption{
Cross-section of the vortex-lattice showing the path taken in a three line
crossing defect. 
\label{fig:6}}
\end{center}
\end{figure}
\begin{figure}[tbp]           
\epsfxsize=10cm
\begin{center}
\leavevmode\epsfbox{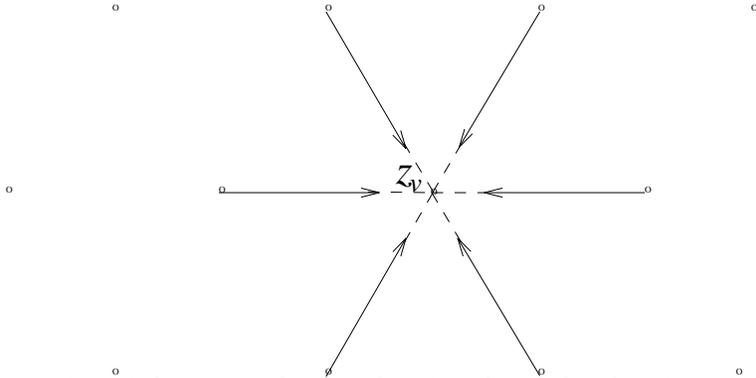}
\caption{
Cross-section of the vortex-lattice showing the path taken in a six line
crossing defect. 
\label{fig:7}}
\end{center}
\end{figure}
The results of the
numerical solutions to the E-L equations are: (i) There is no braid
solution for the three line defect, other than where the three lines meet at
the center of the triangle (as in Fig~\ref{fig:6}). 
(ii) The free energy cost of the 
three lines `crossing' is {\em lower} than the cost of two lines crossing, the
result being:
\begin{equation}
\Delta F_{3\times}=0\cdot 51\;k_BT{|\alpha_T|}^{\frac{3}{2}}.  
\end{equation}
(iii) There is a solution for a six-line braid with an energy cost
of:
\begin{equation}     \label{res6br}
\Delta F_{6\mbox{\scriptsize\em br}}=2\cdot 19\;k_BT{|\alpha_T|}^{\frac{3}{2}}.
\end{equation}
(iv) There is also a solution for the six-line defect where the lines all meet
at a point on the center line (see Fig~\ref{fig:7}),
 and this crossing defect has a
lower free energy cost than the six-line braid:
\begin{equation}                                 
\Delta F_{6\times}=1\cdot 38\;k_BT{|\alpha_T|}^{\frac{3}{2}}.  
\end{equation}

 It will seem surprising that we have found the energy for three lines to meet at
a point to be lower than the energy for two lines. This is due to the potential
term when three lines meet, $c_{000}^{(3)}\simeq 0.54$, being lower than the
potential term for two lines at their midpoint, $c_{00}^{(2)}\simeq 0.79$. Note
that both of these values are for when the surrounding lattice remains fixed,
so it is quite possible that relaxation of the surrounding lattice will reverse
this inequality in the potentials. As was shown in Section~\ref{sec:relax},
allowing the surrounding lines to move may decrease the potential at the
crossing point significantly. However, we also showed in  
Section~\ref{sec:relax} that the full solution including the bending terms in
the free energy does not allow the surrounding lines to relax much, and so we
believe that this result of $\Delta F_{3\times}<\Delta F_{2\times}$ will hold
even when all surrounding lines are taken into account. 

\subsection{Discussion on Two-, Three- and Six-Line Braid Solutions}
 \label{sec:disc}
To compare the calculations for the above defects, and to understand what is
required for braid solutions to exist it is useful to cast the free energies
(\ref{freebr}, \ref{freen}) into the same form. We do this by making the
transformations, $\xi ={(X+iY)}^2$, $\xi= {R_3}^3\exp (i3\theta_3)$, or  
$\xi= {R_6}^6\exp (i6\theta_6)$ respectively. This gives a free energy per
unit length in terms of the complex variable $\xi$:
\begin{eqnarray}  \label{freexi}
f_n\{\xi (h)\}&=&
\sum_{ij=0}^4 d^{(n)}_{ij}
{\left[Re(\xi )\right]}^i {\left[Im(\xi )\right]}^j
+d^{(n)}\left(  {\left( \frac{d \, Re(\xi)}{dh}\right) }^2 +
{\left( \frac{d \, Im(\xi)}{dh}\right) }^2  \right)\\
&=&D_n(\xi)
+d^{(n)}{\left| \frac{d \xi}{dh}\right| }^2 ,\label{freexi2}
\end{eqnarray}
for $n=2,3,6$.
With this substitution, the two- three- and six-line braids all have the same
boundary conditions: $\xi(\pm \infty)=x_n$ where $x_n$ is real, and $\xi$ goes
around the origin of the complex plane once, as $h$ goes from $-\infty$ to
$+\infty$. $x_2=(l/2)^2$, $x_3=(\sqrt{3}l/2)^3$, and $x_6=l^6$. In each case,
$\xi=x_n$ corresponds to the only minimum of $D_n(\xi)$ (the ground state
configuration). In the analogy with classical dynamics, the solutions to these 
boundary conditions which minimize the integral over $h$ of (\ref{freexi2})
correspond to the motion of a particle of mass $m=2d^{(n)}$
in a two-dimensional potential $V(x,y)=-D_n(x+iy)$.
\begin{figure}[htbp]           
\epsfxsize=7cm
\begin{center}
\leavevmode\epsfbox{    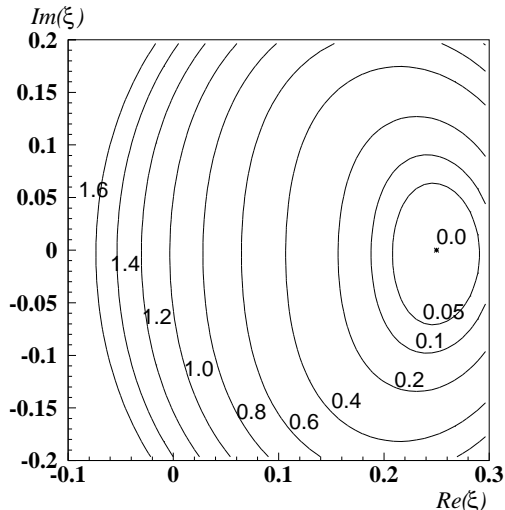}
\caption{ The potential term, $D_2(\xi )=\sum_{i,j} d_{ij}^{(2)}
{\left[ Re(\xi )\right]}^{i}{\left[ Im(\xi )\right]}^{j}$, where $\xi=(X+iY)^2$.
A two-line braid corresponds to starting and ending at the minimum at $(0.25,0)$
while going around the origin once. The path with the lowest potential
will correspond to a crossing defect.
\label{gr:5}}
\end{center}
\end{figure}    
\begin{figure}[htbp]           
\epsfxsize=7cm
\begin{center}
\leavevmode\epsfbox{    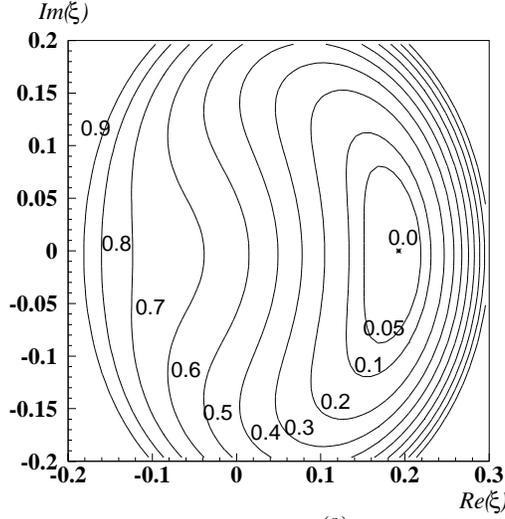}
\caption{   The potential term, $D_3(\xi )=\sum_{i,j} d_{ij}^{(3)}
{\left[ Re(\xi )\right]}^{i}{\left[ Im(\xi )\right]}^{j}$, where
$\xi={R_3}^3\exp{(i3\theta )}$. Again, the three-line braid path with the 
lowest potential will correspond to a crossing defect. 
\label{gr:6}}
\end{center}
\end{figure}
\begin{figure}[htbp]           
\epsfxsize=7cm
\begin{center}
\leavevmode\epsfbox{    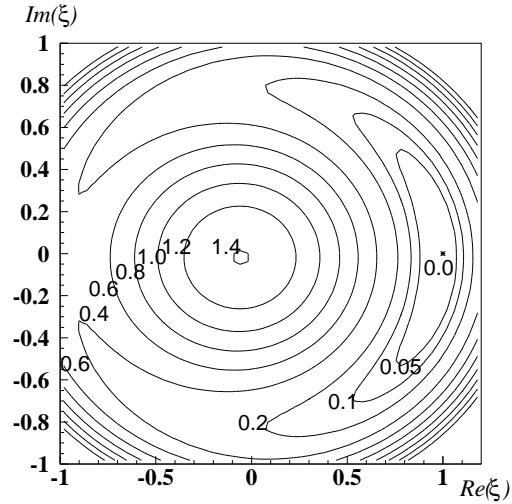}
\caption{   The potential term, $D_6(\xi )=\sum_{i,j} d_{ij}^{(6)}
{\left[ Re(\xi )\right]}^{i}{\left[ Im(\xi )\right]}^{j}$, where
$\xi={R_6}^6\exp{(i6\theta )}$. 
Now there is a `valley path' in the potential that corresponds to a braid
without the lines meeting at the center.
\label{gr:7}}
\end{center}
\end{figure}

It is clear that a braid solution where the lines in the defect
do not meet at their center will only exist if there is a `valley path' in the
$\xi$-plane for the function $D_n(\xi)$ which goes around the origin, starting
and finishing at $\xi=x_n$. Otherwise any braid configuration will be able to 
continuously lower its total free energy by allowing $\xi$ to move nearer
the origin as it goes around and the only
solution  will be  a
crossing defect (where $\xi$ goes to the origin and back along the real axis).
By looking at the forms of $D_n(\xi)$ in 
Figs~\ref{gr:5},~\ref{gr:6}~and~\ref{gr:7} one can see
why there is no braid solution for $n=2,3$, as there are no valley paths.
For $n=6$ the presence of a braid solution
is explained by the almost circular valley path in $D_6$.

The presence of a valley path is a {\em minimum} requirement for a braid
solution to exist. It does not guarantee that this braid will have a
lower energy cost than the corresponding crossing defect, as our results for
$n=6$ show. Even though the height of the potential term $D_6(\xi )$ is 
higher at the origin than at any point in the braid path, the six-line defect
can still lower its total free energy when $\xi$ goes to the origin and back
more than when $\xi$ goes around the valley as the bending term in 
(\ref{freexi2}) will be lower when there is less of a change in
$|\xi|$. For the braid boundary conditions, this
means that for a defect of given length along $h$, the bending term will be
lower when the length of the path in the $\xi$ plane is smaller. Thus there is
a balance here between the potential term which favours the braid defect around
the valley, and the bending term which favours the crossing defect. In this
case the relative terms give the lower total free energy to the crossing defect,
although stationary solutions exist for both defects with a free
energy barrier between the two.

\subsection{The Twelve-Line Braid Defect}

\begin{figure}[htbp]           
\epsfxsize=10cm
\begin{center}
\leavevmode\epsfbox{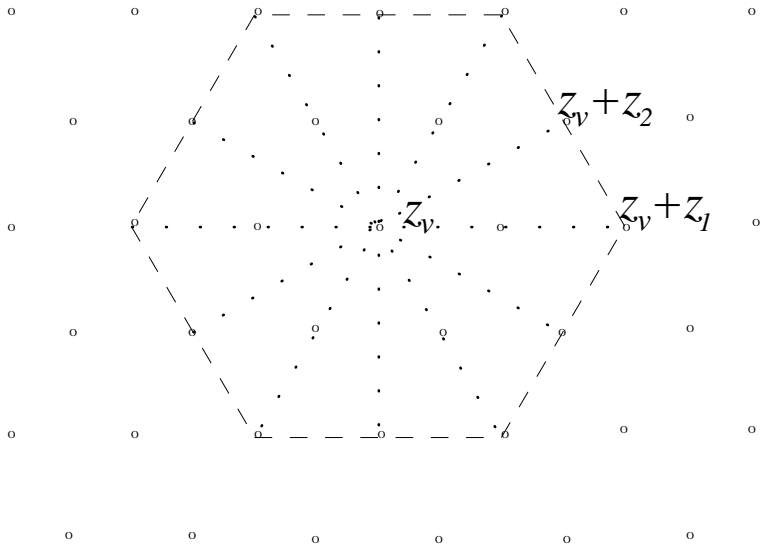}
\vspace{0.5in}\caption{
Cross-section of the vortex-lattice with twelve lines allowed to move within a
defect. The dashed line shows a possible braid path. The dotted line shows the
path of a crossing defect.    
\label{fig:8}}
\end{center}
\end{figure}

 So far in this paper, the symmetry of
the configurations has allowed us to reduce the problems to just two coupled
equations in two parameters, describing the two-dimensional motion of a single
`particle'. In order to consider larger braids than the 2- 3- and 6-line 
defects, one now has to include more degrees of freedom. The simplest problem
we can think of with more than six-lines involved is the `twelve line hexagonal
defect' (see Fig~\ref{fig:8}). 
We can still take advantage of the symmetry of this
configuration by assuming that it always costs least free energy when the six
lines on the vertices of the hexagon move symmetrically about the hexagon's
center, and the six lines on the sides of the hexagon also move symmetrically.
This allows us to reduce the number of degrees of freedom in this defect to
four (corresponding to the two-dimensional motion of two particles in the E-L
equations).

In the ground state, the complex positions of the `edge lines' are given by the
roots of ${(z-z_v)}^6-z_1^6=0$, and the `corner lines' by 
${(z-z_v)}^6-z_2^6=0$, where $z_1=2l$
and $z_2=\sqrt{3}l\exp (i\pi /6)$. If the complex positions relative to $z_v$
move from
$(z_1,z_2)$ to the new positions $(\zeta_1,\zeta_2)$ then the order parameter of
the system can be written:
\begin{equation}
\psi_{12}(x,y,h)=\psi_0(x,y)\frac{({(z-z_v)}^6-{\zeta_1}^6)
({(z-z_v)}^6-{\zeta_2}^6)} 
{({(z-z_v)}^6-{z_1}^6)({(z-z_v)}^6-{z_2}^6)}.
\end{equation}

To find the free energy of twelve-line defects, the same procedure is followed
where we substitute into (\ref{freeint}), then integrate over the $x$-$y$ 
plane, to get a free energy per unit length as a function of the complex 
$\zeta_1(h)$ and $\zeta_2(h)$. We omit the details of the calculation, which
requires a lot of space but results in
four coupled non-linear second order equations in the real and imaginary parts
of ${\zeta_1}^6$ and ${\zeta_2}^6$. This system of four E-L equations was
numerically solved  subject to the braid boundary conditions.
The actual path of the braid of lowest free energy 
was close to the hexagonal perimeter, and this solution gave a total free
energy cost of:
\begin{equation}  \label{res12br}
\Delta F_{12br}\simeq 6\cdot 5\;k_BT{|\alpha_T|}^{\frac{3}{2}}.  
\end{equation}    

A crossing solution was also looked for, where all twelve lines meet at a point
on the central line (at $z=z_v$), moving along the dotted line in
Fig~\ref{fig:8}. For the six line case
this form of defect has a lower energy cost than a braid. 
For twelve lines crossing the result is
\begin{equation}  \label{res12x}
\Delta F_{12\times}\simeq 8\cdot 6\;k_BT{|\alpha_T|}^{\frac{3}{2}}.  
\end{equation}    
This is higher than the twelve line braid free energy. We expect
that all larger braids will have lower energies than the corresponding
crossing defects.

\section{Calculations for Large Defects} \label{sec:large}

In this section, the energy cost is calculated for an infinite straight screw
dislocation (see Fig~\ref{fig:9}), 
and also for two opposite screw dislocations a large distance apart.
 The results of these are used to give the energy cost for very large
braids.  The single screw dislocation may be a first step towards calculating
the energy costs of the screw-edge dislocation loops that are thought to be
important in describing the dynamics of the Abrikosov
crystal\cite{Labusch,Nelson2}. The
creation/growth of large braid defects has been proposed as a mechanism for the
longitudinal resistivity in type-II superconductors \cite{Fisher,Feigel}.

\subsection{Infinite Screw Dislocation} \label{sec:screw}

\begin{figure}[htbp]           
\epsfxsize=15cm
\begin{center}
\leavevmode\epsfbox{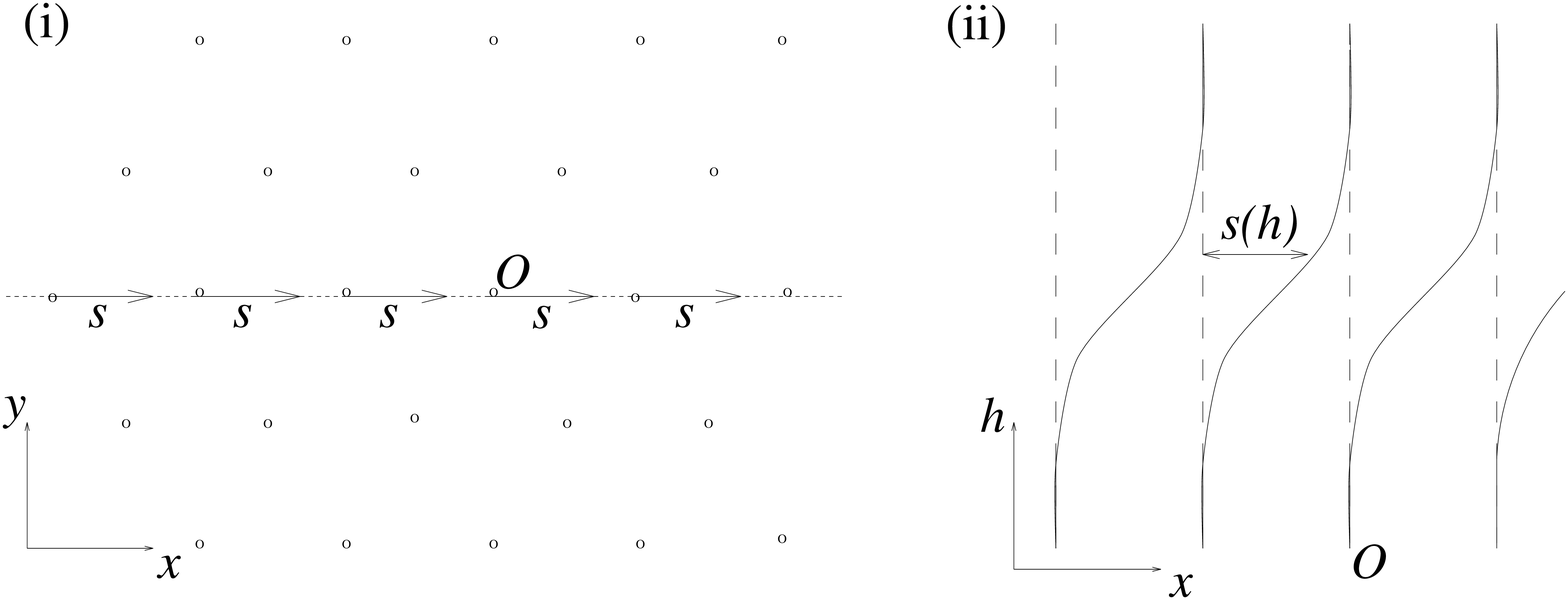}
\caption{
(i) Cross-section of the lattice in the $x$-$y$ plane showing the path of an
infinite straight screw defect. (ii) Schematic side-view of the same defect.  
\label{fig:9}}
\end{center}
\end{figure}

The order parameter of the infinite screw defect in  Fig~\ref{fig:9}
 will depend on only one free parameter, $s(h)$:
\begin{equation}
\psi_{s}(x,y,h)=\psi_0(x,y)\frac{\sin{\pi (z-s(h))}}{\sin{\pi z}},
\end{equation} 
where the origin of the $x$-$y$ plane is now taken to be at the ground state
position of one of the lines involved in this defect. This order parameter
replaces the first order zeros of $\psi_0$ at $z=nl$ with first order zeros at
$z=nl+s(h)$.                                                        

The usual procedure is now followed of integrating the difference of the free
energy density over the $x$-$y$ plane. All of the
integrands are periodic in $x$, giving terms which all diverge linearly with
the system size in the $x$-direction, $L_x$. Also, the bending term contains an
integral which tends to a constant for large $y$, while all other terms are
convergent when integrated over $y$. This leads to the result for the free
energy per unit length as a function of $s(h)$:
\begin{equation}   \label{freescr}
f_s\{ s(h)\} =\sum_{i=1}^2 \frac{L_x}{l} c_i^{(s)} \sin^{2i}\left(\frac{\pi
s}{l}\right)  +\frac{L_xL_y}{l^2} c^{(s)} {\left( \frac{ds}{dh}\right)}^2,
\end{equation}
with $c_1^{(s)}\simeq 0.2883$, $c_2^{(s)}\simeq 2.288\times 10^{-4}$ 
and $c^{(s)}=\pi^2 \langle {\left|\psi_0\right|}^2\rangle \simeq 9.8696$. 
To find the general 
form of $s(h)$ that minimizes the integral over $h$ of (\ref{freescr}), 
for any given system size, we make the substitution 
$h\rightarrow \tilde{h}=h/\xi_s$ with $\xi_s=\sqrt{L_y/l}$. 
This gives a free energy density as a
function of $s(\tilde{h})$ with no dependence on $L_y$ and a linear 
dependence on
$L_x$. The resulting E-L equation in $s(\tilde{h})$ is independent
of $L_x$ and $L_y$,
so increasing  $L_y$ increases the scale
of the screw defect along the $h$-direction, $\xi_s$,  and the resulting 
free energy cost (found by
integrating $f_s\{s(\tilde{h})\}$ over all $\tilde{h}$ 
with $s(\tilde{h})$ the solution to the E-L equation), is
proportional to $L_x\sqrt{L_y}$:
\begin{equation}  
\Delta F_{s}\simeq 0\cdot 54 \frac{L_x\sqrt{L_y}}{l^{3/2}}
\;k_BT{|\alpha_T|}^{\frac{3}{2}}.  
\end{equation}
That is, the free energy cost increases linearly with the number of lines
involved in the defect ($n\propto L_x$) as one would expect, but it also has a 
divergence as the size of the system increases in the direction perpendicular  
to the defect. This is yet another result that is a special feature of the
long-range vortex-vortex interactions in the lowest Landau level, and one that
would not occur with short-range two-body interactions as in the London limit.
The long range dependence of the bending term in the screw defect has important
implications to the free energy cost of large braids, which are developed in
the next section.

\subsection{Two Opposite Screws}

A similar calculation which removes this extra divergence of the energy with
 $L_y$ is of two
`opposite' screws a large distance $W_y=n\sqrt{3}l$ from each other.
This defect has the order parameter:                                     
\begin{equation}
\psi_{2s}(x,y,h)=\psi_0(x,y)\frac{\sin{\pi (z-\frac{nl\sqrt{3}}{2}i-s(h))}
\sin{\pi (z+\frac{nl\sqrt{3}}{2}i+s(h))}}
{\sin{\pi (z-\frac{nl\sqrt{3}}{2}i)}\sin{\pi (z+\frac{nl\sqrt{3}}{2}i)}},
\end{equation} 
where the origin in $(x,y)$ is now at a vortex half-way between the two
opposite screws. For large enough $n$, the potential terms in the free energy
cost for a given $s(h)$ become just twice the corresponding terms for the
single screw. The bending term now contains an integral that diverges
linearly with $n$ rather than $L_y$. The actual result is
\begin{equation} 
f_{2s}\{ s(h)\} =\sum_{i=1}^2 2\frac{L_x}{l} c_i^{(s)} \sin^{2i}\left(\frac{\pi
s}{l}\right)  +4n\sqrt{3}\frac{L_x}{l} c^{(s)} {\left( \frac{ds}{dh}\right)}^2.
\end{equation}

To solve the resulting E-L equation we make a similar transformation as before:
$h\rightarrow \tilde{h}={(2n\sqrt{3})}^{1/2}$. This leads to the same 
E-L equation as for a single screw, 
and the free energy cost of this `double screw' is: 
\begin{equation}  
\Delta F_{2s}\simeq 1\cdot 5 \frac{L_x\sqrt{W_y}}{l^{3/2}}
\;k_BT{|\alpha_T|}^{\frac{3}{2}} .
\end{equation}
As might be deduced from the result for the single screw, 
 the energy of a large double screw is proportional to
the square root of the distance between the two screws. It also suggests
that, at least where the LLL approximation is valid,
the energy cost of large braids will not simply be proportional to the
circumference of the braids (or equivalently the radius of, or the number of
lines in the braid) as has often been suggested. This is shown in the next
section.

\subsection{A Limiting Form For Large Braids}

\begin{figure}[htbp]           
\epsfxsize=10cm
\begin{center}
\leavevmode\epsfbox{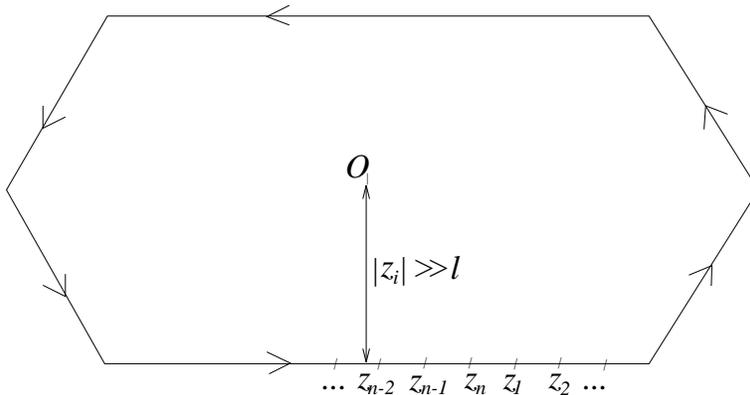}
\caption{
Example of the shape of a very large braid defect in the $x$-$y$ plane where
the vortices move from $z_i$ to $z_{i+1}$ with increasing $h$.
\label{fig:11}}
\end{center}
\end{figure}

We now show that as a braid of general shape becomes very large, the free
energy cost has a simple form depending only on the area enclosed by the braid
and the length of the perimeter of the defect. We consider a large braid
involving $n$ vortices (an example would be the braid in  Fig~\ref{fig:11}), 
where the lines move from $z=z_i$ to $z=z_{i+1}$ as $h$ increases 
between $\pm\infty$, with $i=1,n$ and $z_{n+1}=z_1\exp{(2i\pi)}$. 

Three assumptions are required to
make the necessary approximations: (i) The regions over which the braid is
straight are much larger than regions where the effect of the `corners' of the
braid on the potential term are important. (ii) All lines in the braid move by
the same distance, $s(h)$ towards their nearest neighbor. i.e.\ 
$\zeta_i(h)=z_i+
(z_{i+1}-z_i)(s(h)/l)$. (iii) The braid is everywhere a large distance from the
`center' of the braid. The center, which is where we place the origin, $O$, is
defined by $O=(1/n)\sum z_i$. Therefore this condition means that $|z_i|\gg l$.
These assumptions allow us to approximate the potential term in the free energy
cost using the result for a straight screw (which is proportional to the length
of the screw) and to calculate the bending term in a simple form as follows.

The order parameter of this general $n$-line braid is given by
\begin{equation}
\psi_{n\, br}(x,y,h)=\psi_0(x,y)\frac{\prod_{i=1}^nz-z_i-(z_{i+1}-z_i)(s(h)/l)}
{\prod_{i=1}^nz-z_i},
\end{equation} 
To find the bending term we need the partial derivative of the order parameter 
with respect to $s$:
\begin{eqnarray}
\frac{\partial\psi_{n\, br}}{\partial s}&=&\psi_0
\frac{\sum_{j=1}^n\left\{ -(z_{j+1}-z_j)(1/l)
\prod_{i=1,i\neq j}^n\left( z-z_i-(z_{i+1}-z_i)(s(h)/l)\right)\right\}}
{\prod_{i=1}^n(z-z_i)}\\
&\simeq&
\psi_0
\sum_{j=1}^n\frac{ -(z_{j+1}-z_j)/l}
{z-z_j},
\end{eqnarray} 
where the second line uses condition (iii). We now consider this sum when the
complex coordinates $z=x+iy$ are far from the perimeter of the braid. That is,
 when $|z-z_j| \gg 1$ for all $j$, the sum becomes
\begin{eqnarray}
\frac{\partial\psi_{n\, br}}{\partial s}&=&\sum_{j=1}^n \psi_0 \ln{\left(
\frac{z-z_{j+1}}{z-z_j}\right)}\\
&=&\left\{
\begin{array}{lr}  2i\pi\psi_0 &\mbox{if $z$ inside braid.}\\
0&\mbox{ if $z$ outside braid.}\end{array}\right.
\end{eqnarray}
That is, the partial derivative of the order parameter is periodic within the
braid and zero outside. Although this result does not hold when $z$ is near 
the perimeter, when we
integrate over the $x$-$y$ plane then for large braids the integral will be
dominated by the contribution where this does hold. The limiting result for the
bending term is:
\begin{equation}
\int \frac{d^2r}{l^2}
{\left|\frac{\partial\psi_n}{\partial h}\right|}^2
=\frac{A}{l^2}\pi^2\langle {\left|\psi_0\right|}^2\rangle
{\left(\frac{ds}{dh}\right)}^2, 
\end{equation} 
where ${ A}$ is the area enclosed by the braid. Adding the potential term 
taken from the infinite screw calculation gives a free energy cost for the large
braid as a function of $s(h)$ as:
\begin{equation} 
f_{n\, br}\{ s(h)\} =\sum_{i=1}^2 \frac{P}{l} c_i^{(s)} 
\sin^{2i}\left(\frac{\pi
s}{l}\right)  + 4\frac{{A}}{l^2}c^{(s)}{\left( \frac{ds}{dh}\right)}^2,
\end{equation}
with ${P}$ the length of the perimeter of the braid. Making the correct
transformation in the length scale of the defect, $h\rightarrow
\tilde{h}=h\sqrt{Pl/4A}$ allows us to use  the results
from Section~{\ref{sec:screw}} to find a total free energy cost for a 
large braid of:
\begin{equation}
\Delta F_{n\, br}\simeq 1\cdot 1 \frac{\sqrt{PA}}{l^\frac{3}{2}}
\;k_BT{|\alpha_T|}^{\frac{3}{2}}. \label{reslarge}
\end{equation}

An important caveat for the large defects in this section is that our results
will only be good for defects of less than a certain size.
As in all of our calculations we have used the LLL approximation, we have
assumed that we may ignore fluctuations in the microscopic flux density. 
This will lead to good approximations of the energy costs only when the
size of the defects is less than the distance over which the flux density
 can change
significantly, which is the magnetic penetration depth. In the $x$-$y$ plane,
the dimensions of the defect must be less than $\lambda_{ab}$. The extent of
the defect in the field direction must be less than $\lambda_c$.

\section{Applications} \label{sec:appl}
\subsection{The nature of the vortex state}\label{sec:noent}

The consequences of finding no stable two-line (or three-line) braids, and
of a relatively high free energy cost for the six-line braid greatly affects our
picture of the vortex state in thermal equilibrium at and above the
irreversibility line. If stable two-line braids 
of low free energy existed  they would be created in large numbers by
thermal fluctuations at relatively low temperatures, and give rise to a system
with vortices twisting around each other in the lattice, i.e.\ an entangled
state. However, the absence of such low energy braids means that extensive
entanglement will not occur over a large region of the $H$-$T$ phase diagram
that includes the irreversibility line, as we now show.

We deduce from our results that the braid defect of lowest free energy is the
six-line hexagonal braid, with an energy cost given in Eq.~(\ref{res6br}). 
This is
because we have shown that smaller braids have no saddle point solutions and
the lines can just pull through each other while
 continuously lowering their free
energy. We use the result (\ref{res6br}) to make an estimate of how
many of these braids will be present in a sample of superconducting material of
dimensions typical to those available. The stable six-line braid defect 
extends along the direction of the vortices a distance 
$L_{6\,br}\sim 50\,\xi_c$, where $\xi_c$ is the superconducting
coherence length along the field direction. If we consider a sample with
dimensions along the field axis of $L_c$, then in thermal equilibrium, the
average number of braids along any six lines that surround a given line will be 
$N_{6\,br}=(L_{c}/L_{6\,br})\exp{(-\Delta F_{6\, br}/k_BT)}$. The number of
six-line braids that any one vortex line is involved in will be $6N_{6\,br}$.

If we take an estimate of $\alpha_T$ on the melting line \cite{Wilkin2} 
as $\alpha_T\sim -8$ we find that the exponential gives a factor of  
$\exp{(-\Delta F_{6\, br}/k_BT)}\sim e^{-50}\sim 10^{-22}$. Taking an extreme
lower limit on the coherence length of high-$T_c$ superconductors of
$\xi_c\sim10^{-10}\hbox{m}$, we would need a sample size of
$L_c\simeq10^{13}\hbox{m}$ for there to be an average of one braid defect on
any given line! Alternatively, if we consider the number of six-line braids in
the whole sample, then at a field of 10 T there will be approximately one braid
defect in $10^5\, \hbox{mm}^3$ at the melting line. (Decreasing the external
field reduces the number of defects if we remain on the same $\alpha_T=const$
line).

We now consider at what regimes the braid defects will proliferate. Using the
above estimates we find that, for a sample with $L_c\sim 1\hbox{mm}$ there
will be an average one braid defect per line at $\alpha_T\sim -3$. Above this
line in the $H$-$T$ phase diagram, braid defects will be plentiful and the
vortex state will be strongly entangled. Below this line however, the average
line will not be involved in such defects and the vortices can be well
identified between the top and the bottom of the sample. These arguments make
it clear that below the melting line in the crystalline phase, and for a large
region above in the flux-liquid phase, the vortices in type-II superconductors
of sizes currently available will not be in a strongly entangled state,
when at thermal equilibrium. This is contrary to the picture previously
presented in \cite{Nelson} where entanglement has been claimed to play a major
role in these systems near the irreversibility line. Instead,
the lines maintain a correlation between their positions at the top and and
their positions at the bottom of the sample (for a recent experimental
realization, see \cite{Yao}). Note that this does not mean the lines
are not moving, either by fluctuations transverse to 
their length, or their motion as a whole when in the liquid state. It
simply means that the lines do not twist around each other to form long-lived
topologically distinct entangled states.

Of course, our results are only strictly valid in the high field regime where
the LLL approximation holds. However, if braid defects are of such high energy
costs here, it is hard to see how the Boltzmann factors will come down as we
{\em lower} the temperature or field. Also our calculations are within the
low-temperature crystalline state and it may be questioned if we can extend the
results to the liquid state. These defects depend only locally on the
surrounding lattice, and we expect the liquid state to retain crystalline order
over such length scales for a large region above melting, in which case our
results may still apply. It is also important to note that the absence of
entanglement is a size dependent phenomenon. In a big enough sample,
entanglements will arise in principle, but, at least near the melting line,
they will be completely absent in practice.
The above arguments are only for a system in thermal equilibrium, 
and if the system
has been rapidly cooled from a high temperature to near the melting line  
it is still possible for entanglements to be present for some time.

The absence of any long lived braid defects within the system of
vortices can be tested experimentally. Neutron scattering, or
possibly muon spin resonance, could be used to investigate the 
magnetic fields associated with the braids which are transverse to the
externally applied field.
There should be a locally qualitative difference between the resulting spectra 
 due to the fluctuations in the transverse
displacement of a vortex along its length, and from the local transverse 
field of a stable braid defect.

\subsection{Longitudinal Resistivity due to Braid Defects}

Experiments have been performed where a current is applied between terminals on 
the top and bottom faces of a superconductor, with the external magnetic field
applied parallel to this current \cite{Busch,Safar,Cruz}, allowing measurement
 of the longitudinal resistivity. 
It has been known for some time that the spontaneous production of finite
vortex loops at non-zero temperatures can be a source of non-linear resistivity
in the Meissner-phase of any superconductor \cite{Langer}. 
This idea has been extended to the problem of the resistivity along the
field direction in the mixed state of a type-II superconductor \cite{Feigel}.
If one projects the vortex lines onto the $x$-$y$ plane, then entanglements of
the vortices may be identified as `planar loops' perpendicular to the field
direction. For any current, $J_z$, flowing in this direction, these planar
loops will have an interaction with the current analogous to the vortex loops
in the Meissner-phase. The braid defects that we have calculated are precisely
the sort of defect that project to a planar loop.  

It is important to distinguish between two different limiting
cases. The case we consider is  where the vortex-lattice contains `weak
entanglement' only, i.e.\ entanglements produced by the presence of the current
itself. This should be  distinguished from 
 `strong entanglement' where planar loop type entanglements will exist
at equilibrium, on all length scales, which gives rise to a linear
longitudinal resistivity \cite{Feigel}. 
However, we have shown that in a large region of the phase diagram, there are
very few entanglements present in the size of samples used in experiments, so
that the systems are always in the weak entanglement limit. The fact that the
resistivity has been observed to become non-linear below a line
much higher than the irreversibility line in YBCO \cite{Cruz} 
supports our claims in Section~\ref{sec:noent} concerning the absence of
entanglements.

Following the arguments of Feigelman et.\ al.\cite{Feigel}, but using
our results for the free energy costs of the relevant defects, we find that in
 the presence of a current density $J_z$, a large braid has a total free 
energy cost of:
\begin{equation}
\Delta F({ A}, J_z)= \epsilon_A { A}^{\frac{3}{4}}
-\phi_0 J_z { A},
\end{equation}
with $\epsilon_A\simeq 2.1 \, l^{-3/2}k_BT{|\alpha_T|}^{3/2} 
=2.3 \, (B/\phi_0)^{3/4}k_BT{|\alpha_T|}^{3/2} $. 
Therefore there will be a critical
area, ${ A}_c(J_z)$, above which the braid will be driven to larger growth
by the magnetic interaction with the current:
\begin{equation}
{ A}_c \simeq 8.75\, \frac{ (k_BT)^4 {\alpha_T}^6 B^3}{{\phi_0}^7 {J_z}^4}.
\end{equation}
This will cause dissipation,
which will be proportional to the probability of nucleating large enough braids
by thermal fluctuations. The nucleation of large enough braids requires jumping
a free energy barrier,
which  will lead to a longitudinal resistivity of the form:
\begin{equation} \label{weakres}
\rho_{zz}=\frac{E_z}{J_z}\propto e^{-{\left( \frac{J_T}{J_z}\right) }^3},
\end{equation}
with $J_T\simeq 1.43 \, (B/{\phi_0}^2) k_BT{\alpha_T}^2$. This form is 
quite different to that given previously
\cite{Feigel}, where it was assumed that the free energy cost of planar loops
had a linear dependence on their radius.

Unfortunately, the applicability of the above results to any experiments that
measure this longitudinal resistivity is rather doubtful. This is because at
regions of the $H$-$T$ phase diagram of interest, and for all reasonable
current densities, the critical radius of a braid that needs to be created 
is far larger than the magnetic penetration depth $\lambda_{ab}$, 
(i.e.\ the length scale in the $ab$
plane over which the magnetic field may vary significantly).
Therefore, the important braid defects will be outside the scale where the LLL 
approximation may hold. 

\subsection{Crossing Defects as Energy Barriers to Disentanglement}

The energy barrier $U_{\times}$ to the crossing of two vortices, or to a
process of cutting and reconnection, is the energy cost of the configuration
with the highest free energy that must exist during these processes. This
configuration is assumed to be where the two lines meet each other at a point,
as in the crossing defects that we have calculated. Therefore, our result for
$\Delta F_{2\times}$ for the free energy cost of a two-line crossing defect
maybe used as an estimate of $U_{\times}$. The value of 
$U_{\times}$ has been shown to be an important parameter of the entangled vortex
state \cite{Marchetti,Cates}. The relaxation time of the entangled
vortices grows exponentially with $U_{\times}/k_BT$. A different form for the
longitudinal resistivity to (\ref{weakres}) that applies when there is strong
entanglement\cite{Feigel}  (i.e.\ braids exist on all length scales)
is proportional to $\exp (-U_{\times}/k_BT)$ (as crossing defects will be the
basic energy barrier to the growth of braids). It is doubtful where
this applies though, as we have shown above that the vortex state is not
entangled over a large region of the $H$-$T$ phase diagram.

Estimates of $U_{\times}$ have also been used \cite{Wilkin} 
to explain the transition from
local to non-local resistivity seen in the DC flux transformer experiments on
YBCO \cite{Safar,Cruz}.
In this interpretation, the average length over which crossing defects take
place is associated with a correlation length of vortex `identity'
\cite{Wilkin}. When this length scale becomes smaller than the system size, the
vortices will be uncorrelated between the top and bottom of the sample, and
non-local effects will be of less importance in the resistivity. 
This argument does not depend on the entanglement of the vortices.

\section{Summary}

To summarize, we have calculated the free energy costs of an assortment of
topological defects of the Abrikosov vortex-lattice that is the mean-field
solution for the Ginzburg-Landau free energy functional of a perfect bulk
Type-II superconductor. All of our calculations are made within the lowest
Landau level, which is a good approximation at high fields just below the
$H_{c2}$ line. Although we have made a large approximation in not allowing the
surrounding lattice to move in response to these defects, we  believe that our
answers are still very close to the full solutions with relaxation included.
This is a consequence of the elastic line nature of the lattice, whereby the
vortices' resistance to tilting holds them near to their ground state
positions.

The free energy costs calculated are for crossing defects of 2, 3, 6, and 12
lines; braid defects of 6 and 12 lines (see Table~\ref{tab:res}); infinite screw
defects and a form for very large braids. 
Within the LLL, one cannot truly view the
vortex system in terms of discrete lines with simple two-body interactions
between them (as is the case in the opposite extremes of the London limit).
This non-trivial nature of vortex-vortex interactions has led to a few
unexpected results of our calculations. Firstly there were no braid solutions
for two- or three- line defects -- if any braid configuration of two/three lines
is imposed it may continuously lower its energy by the lines meeting each other
at a point (and thus forming a crossing defect)! Secondly, we have found a
lower free energy cost for three lines crossing than for a two-line crossing.
One assumes that this is a consequence of the lattice symmetry, together with
the effect stated above that the surrounding lines in the lattice do not relax
much in response to a localized defect. Thirdly, it costs a lower free energy for a six-line crossing defect
than a six-line braid, even though the potential terms are higher for the
crossing defect (the tilt terms decrease when the six lines move nearer each
other). This odd result was not carried through to the twelve-line
defects, and we believe that all larger braids cost less energy than the
associated
crossings. Finally, we found that for large screw defects, the free energy cost
diverges with the system size perpendicular to the defect. This effect (a long
range dependence in the tilt energy) leads to a result for
the free energy cost of large braids which depends on the area enclosed by the
braid in a new way.

We think that our results will be useful in interpreting experiments on Type-II
superconductors where the dynamics of the vortices play an important role, at
least in the regimes of high field, and where any layered structure of the
superconductors may be ignored (i.e.\ as long as the coherence length $\xi_c$ is
greater than the inter-layer thickness). Some particular areas of application
were outlined in Section~\ref{sec:appl}, but there are still some
remaining questions yet to be cleared up. For instance, what if any are the
consequences of a lower energy cost of crossing for three-lines than two-lines?
We have not yet been able to use the energy costs calculated to make
quantitative predictions on transport properties such as the resistivity.
However, we have shown that the equilibrium vortex state will not be entangled
over a significant region above and below the irreversibility line of a
high-$T_c$ superconductor. This prediction is open to experimental test,
possibly with a magnetic probe like neutron scattering.

\begin{center}
{\bf ACKNOWLEDGEMENTS}
\end{center}
\bigskip

One of us (MJWD) would like to thank the EPSRC for funding this research.

\appendix
\section{Product Form of The Jacobi-Theta Function}
\label{ap:theta}
In this appendix, it is shown explicitly that $f(z)=\vartheta_3\left(\pi
 z/l,\pi\tau/l\right)$ has the form of (\ref{product}). This
then justifies our procedure in forming a new function with displaced zeros,
but which is still in the LLL subspace, as is done in (\ref{displaced}).
Starting from the standard product representation of a Jacobi-theta function
\cite{Ryzhik}, and for clarity letting $l=1$, we have
\begin{equation}   \label{th3}
\vartheta_3(\pi z,\pi \tau)=\Theta \prod_{n=0}^{\infty} \left( 1-e^
{2\pi i(z-z_n)}\right)\left( 1-e^{-2\pi i(z+z_n)}\right),
\end{equation}
where
$\Theta = \prod_{n=0}^{\infty} \left( 1-e^{2\pi i n\tau}\right) = const$, and
$z_n=-(n+\frac{1}{2})\tau -\frac{1}{2}$.
(For the triangular lattice, $\tau=1/2+i\sqrt{3}/2$ and
$\Theta\simeq 1.0043$.) This is the form that was used to find $\psi_0$ when
numerically calculating the coefficients of Tables~\ref{tab:c2}-\ref{tab:c3}.
 Now, the two products in (\ref{th3}) can be rewritten as
$\left( 1-\exp{[\pm 2\pi i(z\mp z_n)]}\right)= \mp 2i\exp{[\pm \pi i(z\mp z_n)]}
\sin \pi (z\mp z_n)$.
Substituting this expression into (\ref{th3}) gives
\begin{equation}                   \label{th32}
\vartheta_3(\pi z,\pi \tau)=\Theta \prod_{n=0}^{\infty}4e^{-2i\pi z_n}
\sin\pi (z-z_n)\sin\pi (z+z_n).
\end{equation}  
The prefactor, $\prod_{n=0}^{\infty}4\exp{(-2i\pi z_n)}$, can be seen to have
 the correct form to ensure convergence of the product, because as
$n\rightarrow\infty$, for finite $z$, we get:
$\sin \pi(z-z_n) \sin\pi (z+z_n) \simeq (1/4)\exp{(2i\pi z_n)}$.
 Therefore for large $n$, the terms inside the product tend to unity (as is
necessary for an infinite product). Now we can use
the well known product representation of $\sin x=
x\prod_{n=1}^{\infty}\left( 1-({x^2}/{n^2\pi^2}) \right)$.
 Putting the sine terms in (\ref{th32}) in to this form finally reveals the 
product form we are looking for:
\begin{equation}              
\vartheta_3(\pi z,\pi \tau)=\pi^2 \Theta \prod_{n=0}^{\infty}4
e^{-2i\pi z_n}(z+z_n)(z-z_n)
\prod_{i,j=1}^{\infty}\left(1-\frac{{(z+z_n)}^2}{i^2}\right)
\left(1-\frac{{(z-z_n)}^2}{j^2}\right).
\end{equation}
Each of the terms in the second product may be expanded as
$1-{(z\pm z_n)}^2/{j}^2=(1/j^2)\left( j+(z\pm z_n)\right)
\left( j-(z\pm z_n)\right)$. 
This is proportional to the product of two terms of the form $(z-z_j)$, where
the roots, $z_j$, are given by
\begin{eqnarray}
z_j&=&\mp z_n \pm j \nonumber\\
&=&\pm\left( \frac{\tau +1}{2} + n\tau \right)\pm j  .
\end{eqnarray} 
Therefore, we have shown that $f(z)=\vartheta_3\left(\pi
 z/l,\pi\tau/l\right)$ has the general product form of (\ref{product})
and the zeros are positioned periodically in the complex plane, with periods 
$\tau$ and $1$.

\begin{table}
\begin{center}
\caption{\label{tab:c2}Coefficients for two lines allowed to move,
 $c^{(2)}_{ij}$}
\begin{tabular}{cccccc}
$i$ & 0&1&2&3&4\\ \hline 
$c^{(2)}_{i0}$& $0.7905$ & $-5.201$ & $10.0395$ & $-32.7365$ & $100.8234$\\
$c^{(2)}_{i1}$& $5.201$ & $5.1655$ & $-32.7365$ & $403.2937$ & $$\\
$c^{(2)}_{i2}$& $10.0395$ & $32.7365$ & $604.9406$ & $$ & $$\\
$c^{(2)}_{i3}$& $32.7365$ & $403.2937$ & $$ & $$ & $$\\
$c^{(2)}_{i4}$& $100.8234$ & $$ & $$ & $$ & $$\\
\end{tabular}    
\end{center}     

\begin{center}
\caption{\label{tab:c4}Coefficients for two lines crossing with two lines
relaxing, $c^{(4)}_{ij}$, $u_i$, $v_i$, $w_{ij}$}
\begin{tabular}{cccccc}
$i$ & 0&1&2&3&4\\ \hline 
$c^{(4)}_{i0}$& $2.0304$ & $6.2416$ & $8.6774$ & $5.9207$ & $3.3321$\\
$c^{(4)}_{i1}$& $-6.2416$ & $-19.9304$ & $-27.4202$ & $-33.9090$ & $-7.5801$\\
$c^{(4)}_{i2}$& $8.6774$ & $27.4202$ & $40.8076$ & $39.0456$ & $48.8558$\\
$c^{(4)}_{i3}$& $-5.9207$ & $-33.9090$ & $-39.0456$ & $-103.4540$ & $-26.7255$\\
$c^{(4)}_{i4}$& $3.3321$ & $7.5801$ & $48.8558$ & $26.7255$ & $274.8685$\\
\hline 
$u_i$ & $28.7674$ & $-13.9138$ & $28.7674$ & $$ & $$ \\
$v_i$ & $28.7674$ & $13.9138$ & $134.871$ & $$ & $$ \\
$w_{i0}$ & $57.5348$ & $-13.9138$ & $$ & $$ & $$ \\
$w_{i1}$ & $13.9138$ & $269.742$ & $$ & $$ & $$ \\
\end{tabular}    
\end{center}     

\begin{center}
\caption{\label{tab:c3}Coefficients for three and six lines allowed to move,
 $c^{(3)}_{ijk}$,  $c^{(6)}_{ijk}$}
\begin{tabular}{cccccccc}
$ijk$ & 000&200&400&110&220&202&310\\ \hline 
$c^{(3)}_{ijk}$& $0.5443$ & $-23.7064$ & $293.7881$ & $-2.0347$ & $11.6525$
& $11.5790$ & $-15.3239$ \\
$c^{(6)}_{ijk}$& $1.40174$ & $-2.8908$ & $1.32786$ & $-0.18099$ & $0.37540$
& $0.38420$ & $-0.03321$ \\
\end{tabular}    
\end{center}   

\begin{center}
\caption{\label{tab:res}Summary of the Free Energy costs of Crossing and Braid 
Defects}
\begin{tabular}{ccccc}
$i$ & 2&3&6&12\\ \hline 
$\Delta F_{i\times}/(k_BT{|\alpha_T|}^{\frac{3}{2}})$ 
& $0.58$ & $0.51$ & $1.38$ & $8.6$\\
$\Delta F_{i\, br}/(k_BT{|\alpha_T|}^{\frac{3}{2}})$ 
& $$ & $$ & $2.19$ & $6.5$\\
\end{tabular}    
\end{center}      
\end{table}

\end{document}